\documentclass[journal,10pt,onecolumn,twoside]{IEEEtran}%draftcls,onecolumn,

\usepackage{algorithm,algorithmic}
\usepackage{comment}
   \usepackage[pdftex]{graphicx}
  \graphicspath{{Figures/}}
\usepackage{multicol, blindtext}
\usepackage{amsmath,amsthm}
\usepackage{enumerate}
\usepackage{amsfonts}
\title{Kronecker PCA based robust SAR STAP                }

\newtheorem{theorem}{Theorem}[section]

\begin{document}
\author{Kristjan~Greenewald,~\IEEEmembership{Student Member,~IEEE,} Edmund~Zelnio, and~Alfred~Hero III,~\IEEEmembership{Fellow,~IEEE}% <-this % stops a space

\thanks{K. Greenewald and A. Hero III are with the Department
of Electrical Engineering and Computer Science, University of Michigan, Ann Arbor,
MI, USA. E. Zelnio is with the Air Force Research Laboratory, Wright Patterson Air Force Base, OH 45433, USA. This research was partially supported by grants from AFOSR FA8650-07-D-1220-0006 and ARO MURI W911NF-11-1-0391. Approved for public release, PA Approval \#88ABW-2014-6099.
}
}
%

%If tech report
\iftrue
\includecomment{LongerTheorems}
\excludecomment{ShorterTheorems}
\includecomment{ThmApp}
\includecomment{CD}
\includecomment{CD1}
\includecomment{CD2}
\includecomment{CD3}

%Else paper
\else
\excludecomment{LongerTheorems}
\includecomment{ShorterTheorems}
\excludecomment{ThmApp}
\excludecomment{CD}
\excludecomment{CD1}
\excludecomment{CD2}
\excludecomment{CD3}

\fi

 \maketitle
\begin{abstract}
%Say something.

%Kronecker PCA involves the use of a space vs. time Kronecker product decomposition to estimate spatio-temporal covariances. It was shown in \cite{greenewaldArxiv,greenewaldSSP2014} that a diagonally loaded covariance matrix is not well modeled by such a decomposition and that using a diagonal correction factor in the decomposition significantly reduces the required separation rank of the KronPCA estimate. In this work  the addition of a sparse correction factor is considered, which corresponds to a model of the covariance as a sum of Kronecker products and a sparse matrix. This sparse correction includes diagonal correction as a special case but allows for sparse unstructured ``outliers" anywhere in the covariance matrix. This paper introduces a robust PCA-based algorithm to estimate the covariance under this model, extending the nuclear norm penalized LS Kronecker PCA approaches of \cite{greenewaldSSP2014,tsiliArxiv}. This paper also provides an extension to Toeplitz temporal factors, producing a parameter reduction for temporally stationary measurement modeling. High dimensional MSE performance bounds are given for these extensions.
%Finally, the proposed extension of KronPCA is evaluated and compared on both simulated and real data coming from yeast cell cycle experiments. This establishes the practical utility of the sparse correction in biological and other applications.

%SAR GMTI good. Better than GMTI. STAP. KRONPCA WONDERFUL. NEW MODEL. NEW ALGO. NEW PERFORMANCE. REAL DATA. AWESOME RESULTS. EXTENSIONS.

This paper proposes a spatio-temporal decomposition for the detection of moving targets in multiantenna SAR. As a high resolution radar imaging modality, SAR detects and localizes non-moving targets accurately, giving it an advantage over lower resolution GMTI radars. Moving target detection is more challenging due to target smearing and masking by clutter.  Space-time adaptive processing (STAP) is often used to remove the stationary clutter and enhance the moving targets. In this work, it is shown that the performance of STAP can be improved by modeling the clutter covariance as a space vs. time Kronecker product with low rank factors. Based on this model, a low-rank Kronecker product covariance estimation algorithm is proposed, and a novel separable clutter cancelation filter based on the Kronecker covariance estimate is introduced. The proposed method provides orders of magnitude reduction in the required number of training samples, as well as improved robustness to corruption of the training data. Theoretical properties of the proposed estimation algorithm are established showing significant reductions in training complexity under the spherically invariant random vector model (SIRV). Finally, an extension of this approach incorporating multipass data (change detection) is presented. Simulation results and experiments using the Gotcha SAR GMTI challenge dataset are presented that confirm the advantages of our approach relative to existing techniques.

\end{abstract}

%\section{Abstract}
%Goal: Introduce controllable structured sparsity rates thus drastically reducing the number of required samples for multiframe video image covariance estimation.

\section{Introduction}

%Why moving targets. Why radar. Why SAR. Why not MTI.

\IEEEPARstart{T}{h}e detection (and tracking) of moving objects is an important task for scene understanding, as motion often indicates human related activity \cite{newstadt2013moving}. Radar sensors are uniquely suited for this task, as object motion can be discriminated via the Doppler effect. In this work, we propose a spatio-temporal decomposition method of detecting ground based moving objects in airborne Synthetic Aperture Radar (SAR) imagery, also known as SAR GMTI (SAR Ground Moving Target Indication).

Radar moving target detection modalities include MTI radars \cite{newstadt2013moving,ender1999space}, which use a low carrier frequency and high pulse repetition frequency to directly detect Doppler shifts. This approach has significant disadvantages, however, including low spatial resolution, small imaging field of view, and the inability to detect stationary or slowly moving targets. The latter deficiency means that objects that move, stop, and then move are often lost by a tracker.

SAR, on the other hand, typically has extremely high spatial resolution and can be used to image very large areas, e.g. multiple square miles in the Gotcha data collection \cite{GotchaData}. As a result, stationary and slowly moving objects are easily detected and located \cite{ender1999space,newstadt2013moving}. Doppler, however, causes smearing and azimuth displacement of moving objects \cite{jao2001theory}, making them difficult to detect when surrounded by stationary clutter. Increasing the number of pulses (integration time) simply increases the amount of smearing instead of improving detectability \cite{jao2001theory}. Several methods have thus been developed for detecting and potentially refocusing moving targets in clutter. Our goal is to remove the disadvantages of MTI and SAR by combining their strengths (the ability to detect Doppler shifts and high spatial resolution) using space time adaptive processing (STAP) with a novel Kronecker product spatio-temporal covariance model, as explained below.  

%HOW TO USE SAR FOR MOVING TARG DET.

SAR systems can either be single channel (standard single antenna system) or multichannel. Standard approaches for the single channel scenario include autofocusing \cite{fienup2001detecting} and velocity filters. Autofocusing works only in low clutter, however, since it may focus the clutter instead of the moving target \cite{fienup2001detecting,newstadt2013moving}. Velocity filterbank approaches used in track-before-detect processing \cite{jao2001theory} involve searching over a large velocity/acceleration space, which often makes computational complexity excessively high. Attempts to reduce the computational complexity have been proposed, e.g. via compressive sensing based dictionary approaches \cite{khwaja2011applications} and Bayesian inference \cite{newstadt2013moving}, but remain computationally intensive.

Multichannel SAR have the potential for greatly improved moving target detection performance \cite{ender1999space,newstadt2013moving,ginolhac2014exploiting}. Standard multiple channel configurations include spatially separated arrays of antennas, flying multiple passes (change detection), using multiple polarizations, or combinations thereof \cite{newstadt2013moving}. Disadvantages to these approaches include the higher data rate created by collecting multiple channels and the fact that multiple passes involve long delays, registration issues, and having to fly the same orbit more than once \cite{newstadt2013moving}.

%DPCA, ATI, Fienup, STAP. STAP lit review.

\subsection{Previous Multichannel Approaches}
Several techniques exist for using multiple radar channels (antennas) to separate the moving targets from the stationary background. SAR GMTI systems have an antenna configuration such that each antenna transmits and receives from approximately the same location but at slightly different times \cite{GotchaData,ender1999space,newstadt2013moving}. Along track interferometry (ATI) and displaced phase center array (DPCA) are two classical approaches \cite{newstadt2013moving} for detecting moving targets in SAR GMTI data, both of which are applicable only to the two channel scenario. Both ATI and DPCA first form two SAR images, each image formed using the signal from one of the antennas. To detect the moving targets, ATI thresholds the phase difference between the images and DPCA thresholds the magnitude of the difference. A Bayesian approach using a parametric cross channel covariance generalizing ATI/DPCA to $p$ channels was developed in \cite{newstadt2013moving}. Space-time Adaptive Processing (STAP) learns a spatio-temporal covariance from clutter training data, and uses these correlations to filter out the stationary clutter while preserving the moving target returns \cite{ender1999space,ginolhac2014exploiting}. 

A second configuration uses phase coherent processing of the signals output by an antenna array for which each antenna receives spatial reflections of the same transmission at the same time. This contrasts with the above configuration where each antenna receives signals from different transmissions at different times. In this second approach the array is designed such that returns from different angles create different phase differences across the antennas \cite{ginolhac2014exploiting,rangaswamy2004robust,kirsteins1994adaptive,haimovich1996eigencanceler,conte2003exploiting}. In this case, the covariance-based STAP approach, described above, can be applied to cancel the clutter \cite{rangaswamy2004robust,ginolhac2014exploiting,haimovich1996eigencanceler}. 

In this paper, we focus on the first (SAR GMTI) configuration and propose a covariance-based STAP algorithm with a customized Kronecker product covariance structure. The SAR GMTI receiver consists of an array of $p$ phase centers (antennas) processing $q$ pulses in a coherent processing interval. Define the array $\mathbf{X}^{(m)} \in \mathbb{C}^{p\times q}$ such that $X_{ij}^{(m)}$ is the radar return from the $j$th pulse of the $i$th channel in the $m$th range bin. Let $\mathbf{x}_m = \mathrm{vec}(\mathbf{X}^{(m)})$. The radar data $\mathbf{x}_m$ is complex valued and is assumed to have zero mean. Define
\begin{align}
\mathbf{\Sigma} = \mathrm{Cov}[\mathbf{x}] = E[\mathbf{x} \mathbf{x}^H].
\end{align}

The training samples, denoted as the set $\mathcal{S}$, used to estimate the SAR covariance $\mathbf{\Sigma}$ are collected from $n$ representative range bins. %As noted above, the number $pq \choose 2$ of degrees of freedom in the covariance matrix can greatly exceed the number $n = |\mathcal{S}|$ of training samples available to estimate the covariance matrix. This is particularly the case since the  covariance $\mathbf{\Sigma}$ is not constant over all range bins nor over the SAR coherent processing intervals, limiting the number of available training samples. 
The standard sample covariance matrix (SCM) is given by
\begin{align}
\label{Eq:SCM}
\mathbf{S} = \frac{1}{n}\sum_{m\in\mathcal{S}} \mathbf{x}_m \mathbf{x}_m^H.
\end{align}
If $n$ is small, $\mathbf{S}$ may be rank deficient or ill-conditioned \cite{newstadt2013moving,ginolhac2014exploiting,greenewaldArxiv,greenewaldSSP2014}, and it can be shown that using the SCM directly for STAP requires a number $n$ of training samples that is at least twice the dimension $pq$ of $\mathbf{S}$ \cite{reed1974rapid}. In this data rich case, STAP performs well \cite{newstadt2013moving,ender1999space,ginolhac2014exploiting}. However, with $p$ antennas and $q$ time samples (pulses), the dimension $pq$ of the covariance is often very large, making it difficult to obtain a sufficient number of target-free training samples. This so-called ``small $n$ large $p$" problem leads to severe instability and overfitting errors, compromising STAP tracking performance. 

%MORE COMPLETE LIT REVIEW FORTHCOMING.

%SIRV ref: Yao [G 4], also [G 8] 
%SCM needs double the dimension for 3dB loss [G 9]
%Instead of SCM STAP, call LR-STAP [G 10-13].
%SHOW SOME LR-STAP RESULTS ON FEWER PIXELS? FOR BETTER TRAINING SITUATION. OR JUST FEWER PULSES FOR LOWER RANK. IE LOWER PULSE IMAGES. 
%SIRV LR-STAP Performance [G 14]
%SUBSPACE TRACKING [G 15,16]
%STAP ALgos (Krylov, optimized adaptive filters) [G 17-20, 21,22] Robust to outlier [G 23,24]

%SARS are low rank clutter: [N 11], N. Maybe use this to say clutter spatial cov is LR?
%Can geometrically find target on ground [N 7]
%Velocity filterbanks [N 7] [N 12 = compressed sensing dictionary] Prior dist [N].

%For future work better det, ref N.

%????????????????????????????DO A LIT REVIEW, COPY REFS FROM GINOLHAC, OTHERS??????????????????? 

By introducing structure and/or sparsity into the covariance matrix, the number of parameters and the number of samples required to estimate them can be reduced. It has been noted \cite{brennan1992subclutter,ginolhac2014exploiting,ender1999space} that the spatiotemporal clutter covariance $\mathbf{\Sigma}$ is low rank in general, indicating that the clutter lives in a spatiotemporal subspace of dimension $r$. This reduces the number of parameters describing the covariance matrix from $O(p^2q^2)$ to $O(rpq)$. Hence, a common approach to STAP clutter cancelation \cite{ginolhac2014exploiting,rangaswamy2004robust} is to estimate a low rank clutter subspace from $\mathbf{S}$ and use it to estimate and remove the clutter component in the data \cite{bazi2005unsupervised,ginolhac2014exploiting}. We call these methods Low Rank STAP (LR-STAP). Efficient algorithms, including some involving subspace tracking, have been proposed \cite{belkacemi2006fast,shen2009reduced}. Other methods adding structural constraints such as persymmetry \cite{ginolhac2014exploiting,conte2003exploiting}, and robustification to outliers either via exploitation of the SIRV model \cite{ginolhac2009spatio} or adaptive weighting of the training data \cite{gerlach2011robust} have been proposed. Fast approaches based on techniques such as Krylov subspace methods \cite{goldstein1998multistage,honig2002adaptive,pados2007short,scharf2008subspace} and adaptive filtering \cite{rui2011reduced,rui2010reduced} exist. All of these techniques remain sensitive to outlier or moving target corruption of the training data, and generally still require large training sample sizes \cite{newstadt2013moving}. In addition, to the best of our knowledge, none of these techniques explicitly incorporate the known spatio-temporal structure of the data into the covariance estimator. The contribution of this paper is to apply covariance estimation techniques designed to exploit spatio-temporal structure in order to significantly reduce the number $n$ of training samples required as well as to provide a degree of robustness to corrupted training data.

%Many covariance regularization approaches exist for 
%STAP to account for the ``small $n$ large $p$" problem. None of these methods, however, specifically exploit the fact that the STAP covariance matrix has spatio-temporal structure.  The contribution of this paper is to apply covariance estimation techniques designed to exploit spatio-temporal structure in order to significantly reduce the number $n$ of training samples required as well as to provide a degree of robustness to corrupted training data.

%High dimensional covariance. Write down.

%STANDARD STAPS METHOD OF LOW RANKNESS HERE. ALSO PERSYMMETRY, ETC.?

We exploit the explicit space-time arrangement of the covariance by modeling the clutter covariance matrix $\mathbf{\Sigma}_c$ as the Kronecker product of two smaller matrices 
\begin{equation}
\label{KronApprox}
\mathbf{\Sigma}_c = \mathbf{A}\otimes \mathbf{B},
\end{equation}
where $\mathbf{A} \in \mathbb{C}^{p\times p}$ is rank 1 and $\mathbf{B}\in \mathbb{C}^{q\times q}$ is low rank.
%When the measurements are Gaussian with covariance of this form they are said to follow a matrix-normal distribution \cite{tsiligkaridis2013convergence}. This model lends itself to coordinate decompositions \cite{tsiliArxiv}. For spatio-temporal data, we consider the natural decomposition of space (variables) vs. time (frames) \cite{tsiliArxiv,greenewaldArxiv}.
In this setting, the $\mathbf{B}$ matrix is the ``temporal (pulse) covariance" and $\mathbf{A}$ is the ``spatial (antenna) covariance," both determined up to a multiplicative constant. %It is shown that, for the single pass case, the clutter $\mathbf{A}$ is rank one. %An extension to multipass (change detection) GMTI is also presented, for which $\mathbf{A}$ has rank equal to the number of passes.

Kronecker product covariances arise in a variety of applications, including MIMO radar \cite{werner2007estimation}, geostatistics \cite{tsiligkaridis2013convergence}, recommendation systems \cite{allen2010transposable}, multi-task learning \cite{bonilla2008multi}, and genomics \cite{yin2012model}. A rich set of algorithms and associated performance guarantees exist for estimation of covariances in Kronecker product form, including iterative maximum likelihood \cite{werner2008estimation,tsiligkaridis2013convergence}, noniterative L2 based approaches \cite{werner2008estimation}, sparsity promoting methods \cite{tsiligkaridis2013convergence,zhou2014gemini}, and robust ML SIRV based methods \cite{greenewaldSSP2014}. Many of these methods have been shown to achieve significant reductions in the number of training samples required for estimation, in line with the reduction in the number of parameters in the Kronecker covariance model \cite{werner2008estimation,tsiligkaridis2013convergence,zhou2014gemini}. 

%For the SAR GMTI antenna arrangement scheme, the Kronecker representation \eqref{KronApprox} is appropriate for any spatial clutter model \cite{newstadt2013moving,ender1999space} for which the antenna calibration errors can be approximated as constant over the region of interest. In SAR GMTI, this latter condition is satisfied for fairly large regions \cite{newstadt2013moving}. %Hence sufficient training data should always be available for Kronecker covariance estimation in SAR GMTI.

In this paper, an iterative L2 based algorithm is proposed to directly estimate the low rank Kronecker factors from the observed sample covariance. Convergence and symmetric positive semidefiniteness of the estimator is established. Theoretical results indicate significantly fewer training samples are required, and it is shown that the proposed approach improves robustness to corrupted training data. Critically, robustness allows significant numbers of moving targets to remain in the training set. We then introduce the Kron STAP filter, which projects away both the spatial and temporal clutter subspaces. This projects away a higher dimensional subspace than does LR-STAP, thereby achieving improved noise and clutter cancelation. We note that this algorithm differs significantly from the set of methods known as Kron PCA, which involves modeling the covariance as a sum of Kronecker products \cite{tsiliArxiv,greenewaldArxiv,greenewaldTSP}. %Due to the reduction in the number of parameters, a significant reduction in filter estimation variance is achieved.
%Theoretical results demonstrating significant reductions in the number of training samples required are presented.

%DISCUSS ESTIMATION (citations here, one sentence lit review), PARAMETER REDUCTION. SAY STRONGER STRUCTURE THAN SIMPLE LOW RANK. MENTION THE NEW HOSVD TENSOR APPROACH. Also the PF OF LOWER SAMPLES.

%In addition to requiring fewer training samples, it is shown that the proposed approach improves robustness to corrupted training data, critically allowing significant numbers of moving targets to remain in the training set. This is due to the fact that both the clutter and targets have Kronecker covariances with different Kronecker factors. Note that a sum of different Kronecker factors cannot be represented as a single factor model. Hence, in this sense the Kron STAP filter estimate has lower bias than unstructured LR-STAP.
%KronPCA. Standard description and lit review? Brief. Parameter reduction.

%Cancellation estimation

%Shear, multipass, etc.

%We also show that each moving target lives in a one dimensional subspace of Kronecker product form. Hence, once the clutter subspace has been estimated and removed, we can use this fact to detect and focus individual moving targets in a way reminiscent of the single channel focusing approach of \cite{fienup2001detecting}. 

%Discuss new algorithm benefits. Robustness, fewer params, etc.

To summarize, the main contributions of this paper are: 1) the exploitation of the inherent Kronecker product spatio-temporal structure of the clutter covariance; 2) the introduction of the low rank Kronecker product based Kron STAP filter; 3) an algorithm for estimating the spatial and temporal clutter subspaces that is highly robust to outliers due to the additional Kronecker product structure; and 4) theoretical results demonstrating improved signal-to-interference-plus-noise-ratio\begin{CD1}
; and 5) an extension to multipass STAP
\end{CD1}
.

The remainder of the paper is organized as follows. Section \ref{Sec:Model}, presents the multichannel SIRV radar model. Our low rank Kronecker product covariance estimation algorithm and our proposed STAP filter are presented in Section \ref{Sec:KSTAP}\begin{CD2}
with an extension to the case of moving target detection with multiple passes
\end{CD2}
.
Section \ref{Sec:Pred} gives theoretical performance guarantees and Section \ref{Sec:Results} gives simulation results and applies our algorithms to the Gotcha dataset. 

In this work, we denote vectors as lower case bold letters, matrices as upper case bold letters, the complex conjugate as $a^*$, the matrix Hermitian as $\mathbf{A}^H$, and the Hadamard (elementwise) product as $\mathbf{A}\odot \mathbf{B}$.

\section{SIRV Data Model}
\label{Sec:Model}
Let $\mathbf{X}\in \mathbb{C}^{p \times q}$ be an array of radar returns from an observed range bin across $p$ channels and $q$ pulses. We model $\mathbf{x} = \mathrm{vec}(\mathbf{X})$ as a spherically invariant random vector (SIRV) with the following decomposition \cite{yao1973representation,rangaswamy2004robust,ginolhac2014exploiting,ginholhac2013performance}: 
\begin{align}
\label{Eq:decomp}
\mathbf{x} = \mathbf{x}_{target} + \mathbf{x}_{clutter} + \mathbf{x}_{noise} = \mathbf{x}_{target} + \mathbf{n},
\end{align}
where $\mathbf{x}_{noise}$ is Gaussian sensor noise with $\mathrm{Cov}[\mathbf{x}_{noise}]=\sigma^2 \mathbf{I} \in \mathbb{C}^{pq \times pq}$ and we define $\mathbf{n} =\mathbf{x}_{clutter} + \mathbf{x}_{noise}$. The signal of interest $\mathbf{x}_{target}$ is the sum of the spatio-temporal returns from all moving objects, modeled as non-random, in the range bin. The return from the stationary clutter is given by $\mathbf{x}_{clutter}  =\tau \mathbf{c}$ where $\tau$ is a random positive scalar having arbitrary distribution, known as the \emph{texture}, and $\mathbf{c} \in \mathbb{C}^{pq}$ is a multivariate complex Gaussian distributed random vector, known as the \emph{speckle}. We define $\mathrm{Cov}[\mathbf{c}]= \mathbf{\Sigma}_c$ and note that $\mathbf{c}$ is determined by the arrangement of stationary scatters on the ground. The means of these components of $\mathbf{x}$ are zero. The resulting clutter plus noise ($\mathbf{x}_{target} = 0$) covariance is given by
\begin{align}
\label{Eq:Cov}
\mathbf{\Sigma} = E[\mathbf{n}\mathbf{n}^H] =  E[\tau^2]\mathbf{\Sigma}_c + \sigma^2 \mathbf{I}.
\end{align}

The ideal (no calibration errors) random speckle $\mathbf{c}$ is of the form \cite{newstadt2013moving,ender1999space}
\begin{align}
\label{Eq:7}
\mathbf{c} = \mathbf{1}_p \otimes \tilde{\mathbf{c}},
\end{align}
where $\tilde{\mathbf{c}} \in \mathbb{C}^q$. The representation \eqref{Eq:7} follows because the antenna configuration in SAR GMTI is such that each antenna receives signals emitted at different times at the same points in space \cite{newstadt2013moving,GotchaData}. The representation \eqref{Eq:7} gives a clutter covariance of 
\begin{align}
\label{Eq:ClutterCov}
\mathbf{\Sigma}_c = \mathbf{11}^T \otimes \mathbf{B},
\end{align}
where 
\begin{align}
\mathbf{B} = E[\tilde{\mathbf{c}} \tilde{\mathbf{ c }}^H].% =  \mathbf{R}{\mathbf{C}} \mathbf{R}^H,
\end{align}
$\mathbf{B}$ depends linearly on the spatial covariance function ${\mathbf{C}}$ of the clutter reflectivity, which in turn depends on the spatial characteristics of the clutter in the region of interest \cite{ender1999space}. %FIND A REF TO SAY THIS IS OFTEN LOW RANK!!! o.w. just invert.
%While $\mathbf{B}$ is low rank in the case of a scene with a few dominant stationary scatterers on the ground plane, 
While in SAR GMTI $\mathbf{B}$ is not exactly low rank, it is approximately low rank in the sense that significant energy concentration in a few principal components is observed over small regions \cite{borcea2013synthetic}.

%SIGINV HERE?

Due to the long integration time and high cross range resolution associated with SAR, the returns from the general class of moving targets are more complicated. However, if we restrict to targets having constant doppler shift $f$ (proportional to the target radial velocity) within a range bin, the return has the form
\begin{align}
\label{Eq:Moving}
\mathbf{x} = \alpha \mathbf{d} = \alpha \mathbf{a}(f) \otimes \mathbf{b}(f),
\end{align}
where $\alpha$ is the target's amplitude, $\mathbf{a}(f) = [\begin{array}{cccc} 1 & e^{j2\pi \theta_1(f)} & \dots & e^{j \theta_{p}(f)} \end{array}]^T$, the $\theta_i$ depend on doppler shift $f$ and the platform speed and antenna separation \cite{newstadt2013moving}, and $\mathbf{b} \in \mathbb{C}^q$ depends on the target, $f$, and its cross range path. The unit norm vector $\mathbf{d} =  \mathbf{a}(f) \otimes \mathbf{b}(f)$ is known as the \emph{steering vector}. For sufficiently large $\theta_i(f)$, $\mathbf{a}(f)^H \mathbf{1}$ will be small and the target will lie outside of the SAR clutter spatial subspace. Furthermore, as observed in \cite{fienup2001detecting}, for long integration times the return of a moving target is significantly different from that of uniform stationary clutter, implying that moving targets generally lie outside the temporal clutter subspace \cite{fienup2001detecting} as well. %SAY SOMETHING ABOUT B. SHOULD LIE OUTSIDE OF ANY GIVEN STATIONARY SUBSPACE IF SUFFICIENTLY SMALL. HENCE GOOD TO CANCEL IF WILL CANCEL CLUTTER LOTS. OPTIMAL (SORT OF). ANALOGOUS TO SIGINV.

In practice, the signals from each antenna have gain and phase calibration errors that vary slowly across angle and range \cite{newstadt2013moving}. It was shown in \cite{newstadt2013moving} that in SAR GMTI these calibration errors can be accurately modeled as constant over small regions. Let the calibration error on antenna $i$ be $h_i e^{j\phi_i}$ and $\mathbf{h} = [\begin{array}{ccc} h_1 e^{j\phi_1},& \dots, & h_p e^{j\phi_p}\end{array}]$, giving an observed return $\mathbf{x}' = (\mathbf{h} \otimes \mathbf{I})\odot \mathbf{x}$ and a clutter covariance of 
\begin{align}
\label{Eq:KronCov}
\tilde{\mathbf{\Sigma}}_c  = (\mathbf{h} \mathbf{h}^H) \otimes \mathbf{B}= \mathbf{A} \otimes \mathbf{B}
\end{align}
implying that the $\mathbf{A}$ in \eqref{KronApprox} has rank one.

\subsection{Space Time Adaptive Processing}
%Let the vector $\mathbf{d}$ be a spatio-temporal ``steering vector" \cite{ginolhac2014exploiting}, that is, a matched filter for a specific target location/motion profile. The goal of STAP is to use example clutter range bins to train a spatiotemporal filter $\mathbf{F} \in \mathbb{C}^{pq \times pq}$ that cancels the clutter $\mathbf{x}_{clutter}$ (i.e. $\mathbf{d}^H \mathbf{F} \mathbf{x}_{clutter} \approx 0$) while preserving the moving target signals $\mathbf{x}_{target}$ (i.e.  $\mathbf{d}^H \mathbf{F} \mathbf{x}_{target}$ is large).
%\begin{align}
%\label{Eq:6}
%y = \mathbf{d}^H\mathbf{F}\mathbf{x} \approx \mathbf{d}^H\mathbf{x}_{target} + \mathbf{d}^H\mathbf{F}\mathbf{x}_{noise},
%\end{align}
 %For the approximation \eqref{Eq:6} to be achievable requires the target and clutter signals must be separable. DO THIS!!!!!!!

Let the vector $\mathbf{d}$ be a spatio-temporal ``steering vector" \cite{ginolhac2014exploiting}, that is, a matched filter for a specific target location/motion profile. For a measured array output vector $\mathbf x$  define the STAP filter output $y={\mathbf w}^T {\mathbf x}$, where  $\mathbf w$ is a vector of spatio-temporal filter coefficients. By \eqref{Eq:decomp} and \eqref{Eq:Moving} we have
\begin{align}
\label{Eq:Breakdown}
y = \mathbf{w}^H\mathbf{x} = \alpha \mathbf{w}^H \mathbf{d}+ \mathbf{w}^H \mathbf{n}.
\end{align}

The goal of STAP is to design the filter $\mathbf{w}$ such that the clutter is canceled ($\mathbf{w}^H \mathbf{n}$ is small) and the target signal is preserved ($\mathbf{w}^H \mathbf{d}$ is large).
%MAKE THIS CLEARER!! DEFINE N
%In this paper, the filter $\mathbf{w} = \mathbf{F} \mathbf{d}$ .
For a given target with spatio-temporal steering vector $\mathbf{d}$, we say that the filter $\mathbf w$  is an optimal clutter cancellation filter if it maximizes the SINR (signal to interference plus noise ratio), defined as the ratio of the power of the filtered signal $\alpha \mathbf{w}^H \mathbf{d}$ to the power of the filtered clutter and noise \cite{ginolhac2014exploiting}
\begin{align}
\label{Eq:SINRdef}
\mathrm{SINR}_{out} = \frac{|\alpha|^2 |\mathbf{w}^H\mathbf{d}|^2}{E[\mathbf{w}^H \mathbf{n}\mathbf{n}^H \mathbf{w}]} = \frac{|\alpha|^2 |\mathbf{w}^H\mathbf{d}|^2}{\mathbf{w}^H \mathbf{\Sigma} \mathbf{w}},
\end{align}
where $\mathbf{\Sigma}$ is the clutter plus noise covariance in \eqref{Eq:Cov}. 

It can be shown \cite{ender1999space,ginolhac2014exploiting} that, if the clutter covariance is known, under the SIRV model the optimal  filter for targets at steering vector $\mathbf{d}$ is given by the filter
\begin{align}
\mathbf{w} = \mathbf{F}_{opt}\mathbf{d},
\end{align}
where 
\begin{align}
\label{Eq:opt}
\mathbf{F}_{opt} = \mathbf{\Sigma}^{-1}.
\end{align}
Since the true covariance is unknown, we consider filters of the form
\begin{align}
\label{Eq:GenFilt}
\mathbf{w} = \mathbf{F}\mathbf{d},
\end{align}
and use the measurements to learn an estimate of the best $\mathbf F$.
%We will discuss methods and drawbacks of implementing this in practice below.

Traditionally, the sample covariance \eqref{Eq:SCM} has been used to learn the clutter covariance \cite{ginolhac2014exploiting}.
%\begin{align}
%\label{Eq:SCM}
%\mathbf{\Sigma}_{SCM} = \frac{1}{n} \sum_{j=1}^n \mathbf{x}_j \mathbf{x}_j^H.
%\end{align}
%The sample covariance, however, has very noisy low intensity PCA components \cite{mooreSSP2014} when the number of training samples is on the order of the number of variables. As a result, attempting to use the pseudoinverse of the sample covariance is highly unstable and problematic. 
For $n\gg pq$ the inverse of the sample covariance matrix can be used to reliably estimate the optimal filter \eqref{Eq:opt}. Generally, in STAP $n \leq pq$ and a regularized inverse of the sample covariance is often used as an approximation to \eqref{Eq:opt}. For this a dimensionality reduction method called clutter subspace processing can be used, giving an alternative filter $\mathbf{F}$ that approximates \eqref{Eq:opt} by projecting onto a low dimensional subspace. This approach is effective when the clutter subspace is of low rank $r \ll pq$. Most STAP techniques were developed for classical GMTI radars, for which the covariance is low rank by Brennan's rule \cite{brennan1992subclutter}. This approach is also valid for SAR GMTI since the clutter covariance is also low rank \cite{newstadt2013moving,ender1999space}.

%\begin{align}
%A = \Sigma^{-1} = (E[xx^H])^{-1}
%\end{align}

In clutter subspace processing a \emph{clutter subspace} $\{\mathbf{u}_i\}_{i=1}^r$ is estimated using the span of the top $r$ principal components of the clutter sample covariance \cite{ender1999space,ginolhac2014exploiting}. The corresponding clutter cancelation filter is given by the matrix $\mathbf{F}$ that projects onto the space orthogonal to the estimated clutter subspace:
\begin{align}
\label{Eq:RegStap}
\mathbf{F}= \mathbf{I} - \sum_{i=1}^r \mathbf{u}_i \mathbf{u}_i^H.
\end{align}

Since the sample covariance requires a relatively large number of training samples, obtaining sufficient numbers of target free training samples is a practical problem \cite{newstadt2013moving,ginolhac2014exploiting}. In addition, if low amplitude moving targets are accidentally included in training, the sample covariance will be corrupted and partially cancel moving targets as well, which is especially problematic in online STAP implementations \cite{newstadt2013moving,belkacemi2006fast}. The STAP approach discussed below mitigates these problems as it directly takes advantage of the inherent space vs. time Kronecker structure of the clutter covariance $\mathbf{\Sigma}_c$.

\section{Kronecker STAP}
\label{Sec:KSTAP}
\subsection{Kronecker Subspace Estimation}
\label{Sec:Alg}
%FIX !!!!!!!!!!!!! Estimate Kron SIRV cov ML - doable, fast, accurate, but not low rank. If ultra low sample regime, can do, just low rankify a posteriori. In the interest of speed and low ranking, we do L2 based approach a la Werner.

%For the Kronecker product SIRV plus noise covariance model of \eqref{Eq:Cov}, an iterative shrinkage maximum likelihood estimation algorithm was presented in \cite{greenewaldSSP2014}. However this algorithm always gives a full rank estimate \cite{greenewaldSSP2014} and hence does not provide a low rank approximation. To be applicable to STAP, such an approximation is essential in order to identify the clutter subspace. 
%CITE PASCAL SHRINK STUFF, AND SAY WHAT THEY SAY. 
In this section we develop a subspace estimation algorithm that accounts for spatio-temporal covariance structure and has low computational complexity.

Following the approach of \cite{werner2008estimation,greenewaldArxiv,tsiliArxiv,greenewaldSSP2014}, we fit the low rank Kronecker product model \eqref{Eq:KronCov} to the sample covariance matrix ${\mathbf{ S}}$ subject to $\mathrm{rank}(\mathbf{A}) \leq r_a, \mathrm{rank}(\mathbf{B}) \leq r_b$, where the goal is to estimate $E[\tau^2] \mathbf{\Sigma}_c$. The estimation of the parameters $\mathbf{A}$ and $ \mathbf{B}$ in \eqref{Eq:KronCov} is performed by minimizing the following objective function
\begin{equation}
\label{Eq:SparseOpt}
\hat{\mathbf{A}},\hat{\mathbf{B}} = \arg\min_{\mathrm{rank}({\mathbf{A}}) \leq r_a,\mathrm{rank}({\mathbf{B}}) \leq r_b}\| \mathbf{S}-{ \mathbf{A}}\otimes{\mathbf{B}}\|_F^2.
\end{equation}
%which is also closely related to the objective function solved by performing standard PCA on the SCM (i.e. standard LR-STAP).

For a $pq \times pq$ matrix $\mathbf{M}$ define $\{\mathbf{M}(i,j)\}_{i,j=1}^p$ to be its $q \times q$ block submatrices, i.e. $\mathbf{M}(i,j) = [\mathbf{M}]_{(i-1)q + 1:iq,(j-1)q+1:jq}$. Also, let $\overline{\mathbf{M}} = \mathbf{K}_{p,q}^T \mathbf{M} \mathbf{K}_{p,q}$ where $\mathbf{K}_{p,q}$ is the $pq \times pq$ permutation operator such that $\mathbf{K}_{p,q} \mathrm{vec}(\mathbf{N}) = \mathrm{vec}(\mathbf{N}^T)$ for any $p\times q$ matrix $\mathbf{N}$.

The invertible Pitsianis-VanLoan rearrangement operator $\mathcal{R}(\cdot)$ maps $p_tp_s\times p_tp_s$ matrices to $p_t^2 \times p_s^2$ matrices and, as defined in \cite{tsiliArxiv,werner2008estimation} sets the $(i-1)p_t + j$th row of $\mathcal{R}(\mathbf{M})$ equal to $\mathrm{vec}(\mathbf{M}(i,j))^T$, i.e. 
\begin{align}
\label{Eq:SVD}
\mathcal{R}(\mathbf{M}) &= [\begin{array}{ccc} \mathbf{m}_1 & \dots & \mathbf{m}_{p_t^2}\end{array}]^T,\\\nonumber
\mathbf{m}_{(i-1)p_t+j} &= \mathrm{vec}(\mathbf{M}(i,j)), \quad i,j = 1,\dots,p_t.
\end{align}
%If the rank constraints are removed from the objective in \eqref{Eq:SparseOpt}, then by \cite{werner2008estimation,tsiliArxiv,greenewaldArxiv}, 
%\begin{equation}
%\hat{\mathbf{A}}\otimes \hat{\mathbf{B}} = \mathcal{R}^{-1}(\sigma_1 \mathbf{u}_1 \mathbf{v}_1^H),
%\end{equation}
%where $\sigma_1 \mathbf{u}_1 \mathbf{v}_1^H$ is the first singular component of $\mathcal{R}(\mathbf{S})$.

The unconstrained (i.e. $r_a = p, r_b = q$) objective in \eqref{Eq:SparseOpt} is shown in \cite{werner2008estimation,tsiliArxiv,greenewaldArxiv} to be equivalent to a rearranged rank-one approximation problem, with a global minimizer given by
\begin{equation}
\label{Eq:Werner}
\hat{\mathbf{A}}\otimes \hat{\mathbf{B}} = \mathcal{R}^{-1}(\sigma_1 \mathbf{u}_1 \mathbf{v}_1^H),
\end{equation}
where $\sigma_1 \mathbf{u}_1 \mathbf{v}_1^H$ is the first singular component of $\mathcal{R}(\mathbf{S})$. 

When the low rank constraints are introduced, a closed-form solution of \eqref{Eq:SparseOpt} is no longer available. An alternating minimization algorithm is derived in Appendix \ref{App:LRKron} and is summarized by Algorithm \ref{alg:LRKron}. In Algorithm \ref{alg:LRKron}, $\mathrm{EIG}_{r}(\mathbf{M})$ denotes the matrix obtained by truncating the Hermitian matrix $\mathbf M$ to its first $r$ principal components, i.e.
\begin{equation}
\mathrm{EIG}_r(\mathbf{M}) := \sum_{i=1}^r \sigma_i \mathbf{u}_i \mathbf{u}_i^H,
\end{equation}
where $\sum_i \sigma_i \mathbf{u}_i \mathbf{u}_i^H$ is the eigendecomposition of $\mathbf{M}$, and the (real and positive) eigenvalues $\sigma_i$ are indexed in order of decreasing magnitude. The objective \eqref{Eq:SparseOpt} is not convex, but since it is an alternating minimization algorithm, Algorithm \ref{alg:LRKron} gives monotonic convergence of the objective \ref{Eq:SparseOpt} to a local minimum \cite{boyd2009convex}. In practice, we typically initialize LR-Kron with either $\mathrm{EIG}_{r_a}(\hat{\mathbf{A}}),\mathrm{EIG}_{r_b}(\hat{\mathbf{B}})$ where $\hat{\mathbf{A}}, \hat{\mathbf{B}}$ are from the unconstrained estimate \eqref{Eq:Werner}. Monotonic convergence then guarantees that LR-Kron improves on this simple closed form estimator. %INITIALIZATION, ALWAY IMPROVES?

%As THE THING is guaranteed to obtain the global minimum, in practice we often use this estimate to initialize LR-Kron. %Due to monotonic convergence, the resulting estimate is guaranteed to achieve a lower objective value than the initialization.

%The matrices $\mathbf{A}$ and $\mathbf{B}$ are only identifiable up to a scale factor, that is, $\mathbf{A}\otimes \mathbf{B} = (\mathbf{A}/c)\otimes (c \mathbf{B})$ for any scalar $c$. Therefore, in practice it may be advantageous to renormalize them after each iteration so that, for example, $\|\mathbf{A}\|_F = 1$. 
We call Algorithm \ref{alg:LRKron} low rank Kronecker product covariance estimation, or LR-Kron. In Appendix \ref{App:LRKron} it is shown that when the initialization is positive semidefinite Hermitian the LR-Kron estimator $\hat{\mathbf{A}}\otimes \hat{\mathbf{B}}$ is positive semidefinite Hermitian and is thus a valid covariance matrix of rank $r_a r_b$.

\begin{algorithm}[H]
\caption{LR-Kron Covariance Estimation}
\label{alg:LRKron}
\begin{algorithmic}[1]
%\STATE $\mathbf{B} = \mathbf{P}\mathcal{R}(\hat{\mathbf{\Sigma}}_{SCM})$
\STATE $\mathbf{S} = \mathbf{\Sigma}_{SCM}$, form $\mathbf{S}(i,j)$, $\overline{\mathbf{S}}(i,j)$.
\STATE Initialize $\mathbf{A}$ s.t. $\|\mathbf{A}\|_F = 1$ (or correspondingly $\mathbf{B}$).
\WHILE{Objective $\| \mathbf{S}-{ \mathbf{A}}\otimes{\mathbf{B}}\|_F^2$ not converged}%Objective \eqref{Eq:SparseOpt} not converged}
\STATE $\mathbf{R}_B= \frac{\sum_{i,j}^p a^*_{ij}\overline{\mathbf{S}}(i,j)}{\|\mathbf{A}\|_F^2}$
\STATE ${\mathbf{B}} = \mathrm{EIG}_{r_b}(\mathbf{R}_B)$
\STATE $\mathbf{R}_A= \frac{\sum_{i,j}^q b^*_{ij}\mathbf{S}(i,j)}{\|\mathbf{B}\|_F^2}$
\STATE $\mathbf{A} = \mathrm{EIG}_{r_a}(\mathbf{R}_A)$
\ENDWHILE
\RETURN $\hat{\mathbf{A}} = {\mathbf{A}},\hat{\mathbf{B}} = {\mathbf{B}}$.
\end{algorithmic}
\end{algorithm}

%The following corollary is proven in Appendix \ref{App:LRKron}:
%\begin{corollary}
%\label{Cor:PSD}
%The LR-Kron estimator $\hat{\mathbf{A}}\otimes \hat{\mathbf{B}}$ given by Algorithm \ref{alg:LRKron} (for finite or infinite iterations) is positive semidefinite Hermitian and is thus a valid covariance matrix.
%\end{corollary}

%TRY TO SHOW NO LOCAL MINIMA. 

\subsection{Robustness Benefits}
\label{Sec:Robust}
Besides reducing the number of parameters, Kronecker STAP enjoys several other benefits arising from associated properties of the estimation objective \eqref{Eq:SparseOpt}.

%\subsubsection{Standard PCA Shortcomings}
The clutter covariance model \eqref{Eq:KronCov} is low rank, motivating the PCA singular value thresholding approach of classical STAP. This approach, however, is problematic in the Kronecker case because of the way low rank Kronecker factors combine. Specifically, the Kronecker product $\mathbf{A}\otimes \mathbf{B}$ has the SVD \cite{loan1992approximation}
\begin{align}
\mathbf{A}\otimes \mathbf{B}=(\mathbf{U}_B\otimes \mathbf{U}_B) (\mathbf{S}_A\otimes \mathbf{S}_B) (\mathbf{U}_A^H\otimes \mathbf{U}_B^H)
\end{align}
where $\mathbf{A}= \mathbf{U}_A \mathbf{S}_A \mathbf{U}_A^H$ and $\mathbf{B} = \mathbf{U}_B \mathbf{S}_B \mathbf{U}_B^H$ are the SVDs of $\mathbf{A}$ and $\mathbf{B}$ respectively. The singular values are $s_A^{(i)} s_B^{(j) }, \: \forall i,j $.
As a result, a simple thresholding of singular values is not equivalent to separate thresholding of the singular values of $\mathbf{A}$ and $\mathbf{B}$ and hence won'€™t necessarily adhere to the space vs. time structure.% (see Figure \ref{Fig:KronSpectrum}).

For example, suppose that the set of training data is corrupted by inclusion of a sparse set of $w$ moving targets. By the model \eqref{Eq:Moving}, the $i$th moving target gives a return (in the appropriate range bin) of the form
\begin{equation}
\mathbf{z}_i  = \alpha_i \mathbf{a}_i \otimes \mathbf{b}_i,
\end{equation}
where $\mathbf{a}_i,\mathbf{b}_i$ are unit norm vectors. %The $\mathbf{z}_i$ additively corrupt the range bins $m_i$ in the training data.

This results in a sample data covariance created from a set of observations $\mathbf{n}_m$ with $\mathrm{Cov}[\mathbf{n}_m]= \mathbf{\Sigma}$, corrupted by the addition of a set of $w$ rank one terms
\begin{align}
\label{Eq:corrupt}
\mathbf{S} = \left(\frac{1}{n} \sum_{m=1}^{n} \mathbf{n}_m \mathbf{n}_m^H \right) + \frac{1}{n} \sum_{i = 1}^{w} \mathbf{z}_i \mathbf{z}_i^H .
%\mathbf{\Sigma} = \mathbf{A}\otimes \mathbf{B} + \sum_{i=1}^w \lambda_i \mathbf{u}_i \mathbf{v}_i^H.
\end{align}
%where the latter approximation holds when $\sigma \rightarrow 0$ and the moving targets are indeed orthogonal to the clutter subspace.
%where $\mathrm{Cov}[\mathbf{x}_m] = \mathbf{\Sigma}$.
%\begin{equation}
%E[\mathbf{x}_m \mathbf{x}_m^H] = \mathbf{\Sigma} =  E[\tau^2]\mathbf{A}\otimes \mathbf{B} + \sigma^2 \mathbf{I}.
%\end{equation}

%Since the reflectivity of manmade moving targets is high, $|\alpha_i|$ is often large. 
Let $\tilde{\mathbf{S}} = \frac{1}{n} \sum_{m=1}^{n} \mathbf{n}_m \mathbf{n}_m^H$ and $\tilde{\mathbf{T}} = \frac{1}{n} \sum_{i = 1}^{w} \mathbf{z}_i \mathbf{z}_i^H$. Let ${\lambda}_{S,k}$ be the eigenvalues of $\mathbf{\Sigma}_c$, $\lambda_{S,min} = \min_{k} \lambda_{S,k}$, and let $\lambda_{T,max}$ be the maximum eigenvalue of $\tilde{\mathbf{T}}$. Assume that moving targets are indeed in a subspace orthogonal to the clutter subspace. %When for any $i$, $\frac{1}{n} |\alpha_i|^2 > O(\hat{\lambda}_r)$, performing rank $r$ PCA on $\mathbf{S}$ 
If $\lambda_{T,max} > O(\lambda_{S,min})$, performing rank $r$ PCA on $\mathbf{S}$ 
will result in principal components of the moving target term being included in the ``clutter" covariance estimate. %This of course assumes that moving targets are indeed in an approximately orthogonal subspace to the clutter.

If the targets are approximately orthogonal to each other (i.e. not coordinated), then $\lambda_{T,max} = O(\frac{1}{n} |\alpha_i|^2)$. Since the smallest eigenvalue of $\mathbf{\Sigma}_c$ is often small, this is the primary reason that classical LR-STAP is susceptible to moving targets in the training data \cite{newstadt2013moving,ginolhac2014exploiting}. 

On the other hand, Kron-STAP is significantly more robust to such corruption. Specifically, consider the \emph{rearranged} corrupted sample covariance:
%REQUIRES EXPLANATION OF REARRANGEMENT OPERATOR.
\begin{equation}
\mathcal{R}(\mathbf{S}) = \frac{1}{n} \sum_{m=1}^w \mathrm{vec}(\mathbf{a}_i\mathbf{a}_i^H) \mathrm{vec}( \mathbf{b}_i \mathbf{b}_i^H)^H + \mathcal{R}(\tilde{\mathbf{S}}).
\end{equation}
This also takes the form of a desired sample covariance plus a set of rank one terms. For simplicity, we ignore the rank constraints in the LR-Kron estimator, in which case we have \eqref{Eq:Werner}
\begin{equation}
\hat{\mathbf{A}}\otimes \hat{\mathbf{B}} = \mathcal{R}^{-1}(\hat{\sigma}_1 \mathbf{u}_1 \mathbf{v}_1^H),
\end{equation}
where $\hat{\sigma}_1 \mathbf{u}_1 \mathbf{v}_1^H$ is the first singular component of $\mathcal{R}(\mathbf{S})$. Let ${\sigma}_1$ be the largest singular value of $\mathcal{R}(\tilde{\mathbf{S}})$. %Then, using KronPCA \eqref{Eq:SVD}, the Kronecker product covariance estimate will select 
The largest singular value $\hat{\sigma}_1$ will correspond to the moving target term only if the largest singular value of $\frac{1}{n} \sum_{m=1}^w \mathrm{vec}(\mathbf{a}_i\mathbf{a}_i^H) \mathrm{vec}( \mathbf{b}_i \mathbf{b}_i^H)^H$ is greater than $O(\sigma_1)$. If the moving targets are uncoordinated, this holds if for some $i$, $\frac{1}{n} |\alpha_i|^2  > O({\sigma}_1)$.
Since $\sigma_1$ models the entire clutter covariance, it is on the order of the total clutter energy, i.e. $\sigma_1^2 = O(\sum_{k=1}^r \lambda_{S,k}^2) \gg \lambda_{S,min}^2$. In this sense Kron-STAP is much more robust to moving targets in training than is LR-STAP. %Both methods are of course still susceptible to singular vector corruption. %Also note that our Kron-STAP algorithm imposes additional structure in the form of low rank Kronecker factors \eqref{Eq:SparseOpt}, which provides additional protection against corruption of the covariance estimate.

\subsection{Kronecker STAP Filters}
\label{Sec:STAP}
%\subsubsection{KASSPER}
%\subsection{Filters}

Once the low rank Kronecker clutter covariance has been estimated using Algorithm \ref{alg:LRKron}, it remains to identify a filter $\mathbf F$, analogous to \eqref{Eq:RegStap}, that uses the estimated Kronecker covariance model. If we restrict ourselves to subspace projection filters and make the common assumption that the target component in \eqref{Eq:decomp} is orthogonal to the true clutter subspace, then the optimal approach in terms of SINR is to project away the clutter subspace, along with any other subspaces in which targets are not present. If only target orthogonality to the joint spatio-temporal clutter subspace is assumed, then the optimal STAP filter is the projection matrix: %Based on \eqref{Eq:opt}, if as is usually assumed \cite{ginolhac2014exploiting} the (moving) targets of interest do not lie in the clutter subspace, the appropriate filter is thus

%Under the model (6), under the common assumption that  the target component is orthogonal to the clutter subspace \cite{ginolhac2014exploiting}, then the minimum mean squared error estimator of the target is the projection of the observations onto the clutter subspace. Thus the optimal STAP filter is the projection matrix:

\begin{align}
\label{Eq:KronUS}
\mathbf{F}_{classical} = \mathbf{I} - \mathbf{U}_A \mathbf{U}_A^H\otimes {\mathbf{U}_B \mathbf{U}_B^H},
\end{align}
where $\mathbf{U}_A, \mathbf{U}_B$ are orthogonal bases for the rank $r_a$ and $r_b$ subspaces of the low rank estimates of $\mathbf{A}$ and $\mathbf{B}$, respectively, obtained by applying Algorithm \ref{alg:LRKron}. This is the Kronecker product equivalent of the standard STAP projector \eqref{Eq:RegStap}. %We note that in practice it is rarely the case that moving target signals are fully orthogonal to the true clutter subspace, however they should be approximately so \cite{ginolhac2014exploiting}.

Additional information is available, however. Specifically, by \eqref{Eq:Moving}, no moving target should lie in the same spatial subspace as the clutter. We thus propose the spatial-only filter (which we call spatial-only Kron STAP)
\begin{align}
\label{Eq:SPKR}
\mathbf{F}_{spatial} = (\mathbf{I}-\mathbf{U}_A \mathbf{U}_A^H)\otimes \mathbf{I}.
\end{align}
to cancel as much of the clutter ``subspace leakage" as possible while minimizing target cancelation. This leakage is due to noise and covariance estimation errors. Furthermore, as noted in Section \ref{Sec:Model}, if the dimension of the clutter temporal subspace is sufficiently small relative to the dimension $q$ of the entire temporal space, moving targets will have temporal factors ($\mathbf{b}$) whose projection onto the clutter temporal subspace are small. Under these assumptions, it is thus near-optimal to project away both the temporal and spatial clutter subspaces. We thus propose a Kronecker STAP filter $\mathbf{F}_{KSTAP}$ of the following form:
\begin{align}
\label{Eq:KronSTAP}
\mathbf{F}_{KSTAP} = (\mathbf{I} - \mathbf{U}_A \mathbf{U}_A^H) \otimes (\mathbf{I} - \mathbf{U}_B \mathbf{U}_B^H) = \mathbf{F}_A \otimes \mathbf{F}_B.
\end{align}
We denote by Kron-STAP the method using LR-Kron to estimate the covariance and \eqref{Eq:KronSTAP} to filter the data. Our clutter model has spatial factor rank $r_a = 1$ \eqref{Eq:KronCov}, implying that the $\mathbf{F}_{KSTAP}$ defined in \eqref{Eq:KronSTAP} projects the array signal $\mathbf x$ onto a $(p-1)(q-r_b)$ dimensional subspace. This is significantly smaller than the $pq-r_b$ dimensional subspace onto which \eqref{Eq:KronUS} and unstructured STAP project the data. As a result, much more of the clutter that ``leaks" outside the primary subspace can be canceled, thus allowing lower amplitude moving targets to be detected. %We call the filter of \eqref{Eq:KronSTAP} Kronecker STAP (KronSTAP).

\begin{CD}
\subsection{Multipass STAP}
\label{Sec:MultiPass}
In surveillance applications, it is often of interest to determine what, if anything, has changed in a scene between a reference time $t_0$ and a later time $t_1$, e.g. disappearance/appearance of parked vehicles, or the appearance of vehicle footprints \cite{newstadt2013moving,bazi2005unsupervised,bovolo2005detail,ranney2006signal}. When SAR is used for such change detection applications, the radar platform will generally fly past the scene and form a ``reference'' image at time $t_0$, and then at time $t_1 > t_0$ fly a path as close as possible to the original and form a new ``mission'' image. These images are then compared and changes detected. However, moving targets will almost always be detected as changes, along with the changes in the stationary scene background \cite{newstadt2013moving}. When changes of background are of primary interest, moving targets may in fact mask changes in the stationary scene due to displacement and smearing. Hence, it is advantageous to identify moving targets in both scenes prior to or parallel to background change detection. In addition, it may be of interest to detect moving targets in the imagery for their own sake \cite{newstadt2013moving}. We thus exploit the additional scene information arising from having two images to better estimate the clutter subspace, and follow STAP with subsequent noncoherent change detection. %The latter is possible because any given moving target can only appear in one of the two images at the same location and speed, due to the time difference.

Our Kronecker STAP based change detection approach concatenates the spatial channels of both registered phase histories ($\mathbf{X}_k$), forming a ``$2p$ channel phase history"
\begin{align}
\mathbf{X} = \left[\begin{array}{c}\mathbf{X}_1\\ \mathbf{X}_2\end{array}\right] \in \mathbb{C}^{2p \times q}.
\end{align}
Since two images are involved with potentially different calibration errors, the clutter subspace is of rank 2. Thus, a rank 2 spatial clutter subspace and a low rank temporal subspace are estimated using LRKron and projected away via the KronSTAP filter. This two pass procedure is easily extended to handle multiple ($>2$) passes of the radar sensor.

%Finally, noncoherent change detection is performed by reforming each image and subtracting the resulting pixel magnitudes. %IS THIS TRUE??  %As in the single pass case, the amount of cancelation can be computed and thresholded.

\end{CD}

%%%%%%%%%%%%%%%%%%%%%%%%%%%%%%%%%%%%%%%%%%%%%%%%%%%%%%%%%%%%%%%%%%%%%%%%%%%%%%%%%%%%%%
\begin{LongerTheorems}
\section{SINR Analysis}
\label{Sec:Pred}
%USE CONCENTRATION IN TOP SINGS AS PARAM. FOR ORIGINAL AMT OF CANCEL.
%*Follow the approach of \cite{ginolhac2014exploiting}.
%**Analyzes the effect of noise on subspace estimation error for large n (quadratic perturbation analysis).
%**Converts to a ratio between the achieved SINR and that achievable with infinite training.

For a STAP filter matrix $\mathbf{F}$ and steering vector $\mathbf{d}$, the data filter vector is \eqref{Eq:GenFilt} $\mathbf{w} = \mathbf{F}\mathbf{d}$ \cite{ginolhac2014exploiting}. With a target return of the form $\mathbf{x}_{target} = \alpha \mathbf{d}$, the filter output is given by \eqref{Eq:Breakdown}, and the SINR by \eqref{Eq:SINRdef}.

%\begin{align}
%y = \mathbf{w}^H\mathbf{x} = \alpha \mathbf{w}^H \mathbf{d}+ \mathbf{w}^H \mathbf{n}.
%\end{align}
%MAKE THIS CLEARER!! DEFINE N
%In this paper, the filter $\mathbf{w} = \mathbf{F} \mathbf{d}$ .

%The SINR at the steering vector $\mathbf d$ \cite{ginolhac2014exploiting} is then given by
%\begin{align}
%\mathrm{SINR}_{out} = \frac{|\alpha|^2 |\mathbf{w}^H\mathbf{d}|^2}{E[\mathbf{w}^H \mathbf{n}\mathbf{n}^H \mathbf{w}]} = \frac{|\alpha|^2 |\mathbf{w}^H\mathbf{d}|^2}{\mathbf{w}^H \mathbf{\Sigma} \mathbf{w}}.
%\end{align}
%where $\mathbf{\Sigma}$ is the clutter plus noise covariance in \eqref{Eq:Cov}.

Define $\mathrm{SINR}_{max}$ to be the optimal SINR, achieved at $\mathbf{w}_{opt} = \mathbf{F}_{opt} \mathbf{d}$ \eqref{Eq:opt}.

Suppose that the clutter has covariance of the form \eqref{Eq:KronCov}. %, i.e. it lies in an $r_b$ dimensional subspace, but some ``subspace leakage" occurs. Define ? and ? to be the ratios of ?. 
Assume that the target steering vector $\mathbf{d}$ lies outside both the temporal and spatial clutter subspaces as per above and \cite{ginolhac2014exploiting}. Suppose that LR-STAP is set to use $r$ principal components. Suppose further that Kron STAP uses 1 spatial principal component and $r$ temporal components, so that the total number of principal components of LR-STAP and Kron STAP are equivalent. 

Under these assumptions, if $\sigma$ approaches zero the SINR achieved using LR-STAP, Kron STAP or spatial Kron STAP with infinite training samples achieves \cite{ginolhac2014exploiting} $\mathrm{SINR}_{max}$. %$SINR_{max} = \frac{|\tau|^2}{\sigma^2}$ and achieved at some $\mathbf{w}_{opt}$.
%Similarly, the SINR achieved using KronSTAP with infinite training samples is given by
%\begin{align}
%\end{align}.
%Already a gain. OR IS IT? ONLY IF SUBSPACE LEAKAGE INTO SUBSPACES NOT QUITE ORTHOGONAL TO STEERING VECTOR!!! THEN IT IS. HENCE, SAY EACH STAGE ALONE, UNDER THESE ASSUMPTIONS, IS EQUAL. COMBINING STAGES ONLY HELPS IN AFOREMENTIONED SETTING.

We analyze the asymptotic convergence rates under the finite sample regime. Define the SINR Loss $\rho$ as the loss of performance when using the estimate $\hat{\mathbf{w}} = \hat{\mathbf{F}}\mathbf{d}$ (corresponding to $\mathrm{SINR}_{out}$) as the filter instead of $\mathbf{w}_{opt}$:
\begin{align}
\rho = \frac{\mathrm{SINR}_{out}}{\mathrm{SINR}_{max}}.
\end{align}

Let $\lambda_i$, $i = 1,\dots,pq$ be the eigenvalues of $\mathbf{\Sigma}_c$. Under the Kronecker model, we have
\begin{equation}
{\lambda}_i = \left\{\begin{array}{ll}s_A^{(1)} s_B^{(i)}, & i = 1,\dots, r_b \\ 0 & i > r_b\end{array}\right.
\end{equation}
%and $\lambda_i = 0$ for $i > q$, 
since $\mathbf{A}$ only has one nonzero singular value.

\begin{theorem}[LR-STAP SINR \cite{ginolhac2014exploiting}]
\label{Thm:LRSTAP}
%*Degradation of target/clutter SINR (LR-STAP):

For large $n$, the expected SINR Loss of LR-STAP is

\begin{align}
E[\rho] = 1-\frac{1}{n}\sum_{i=1}^r\left(\frac{E[\tau^2]\lambda_i + \sigma^2}{E[\tau^2]\lambda_i}\right)^2,
\end{align}
which in the small $\sigma^2$ regime (typical in SAR \cite{ginolhac2014exploiting}) becomes
\begin{align}
E[\rho] \approx 1-\frac{r}{n}
\end{align}
\end{theorem}
Under the Kronecker model we have
\begin{align}
E[\rho] = 1-\frac{1}{n}\sum_{i=1}^{r}\left(\frac{E[\tau^2]s_B^{(i)} + \frac{\sigma^2}{s_A^{(1)}}}{E[\tau^2]s_B^{(i)}}\right)^2.
\end{align}

We now turn to Kron STAP. Note that the Kron STAP filter can be decomposed into a spatial stage (filtering by $\mathbf{F}_{spatial}$) and a temporal stage (filtering by $\mathbf{F}_{temp}$):
\begin{equation}
\mathbf{F}_{KSTAP} = \mathbf{F}_A\otimes \mathbf{F}_B = \mathbf{F}_{spatial} \mathbf{F}_{temp}
\end{equation}
where $\mathbf{F}_{spatial} = \mathbf{F}_A \otimes \mathbf{I}$ and $\mathbf{F}_{temp} = \mathbf{I} \otimes \mathbf{F}_B$ \eqref{Eq:KronSTAP}.
Under the idealized model in this section, either the spatial or the temporal stage is sufficient to project away the clutter subspace. We assume the naive estimator 
\begin{align}
\label{Eq:Spat}
\hat{\mathbf{A}} = \mathrm{EIG}_1\left(\frac{1}{q}\sum_{i} \mathbf{S}(i,i) \right) = \hat{\psi}\hat{\mathbf{h}}\hat{\mathbf{h}}^H
\end{align}
for the spatial subspace $\mathbf{h}$ ($\|\mathbf{h}\|_2=1$). This is equivalent to approximating the sample spatial covariance as rank 1. The analysis of \cite{ginolhac2014exploiting} thus applies with $r=1$ and $n' = nq$, except some of the samples are correlated. Using the Kronecker structure of the covariance it is trivial to show (for the SIRV distribution) that the worst case occurs when all the clutter temporal correlations are all $\pm1$, in which case $\frac{1}{q} \sum_i \mathbf{S}(i,i)$ reduces to an $n$ iid sample SCM with Gaussian noise variance $\sigma^2/q$ and we can directly obtain the following via Theorem \ref{Thm:LRSTAP}
\begin{theorem}[Kron STAP SINR]
\label{Thm:Sp}
For large $n$ and using the estimator \eqref{Eq:Spat}, the expected SINR Loss of Kron STAP using the estimator \eqref{Eq:Spat} for the spatial subspace satisfies
\begin{align}
E[\rho] \geq 1-\frac{1}{n}\left(\frac{E[\tau^2]\psi + \frac{\sigma^2}{q}}{E[\tau^2]\psi}\right)^2 %\leq  \leq 1-\frac{1}{qn}\left(\frac{E[\tau]\phi_1 + \rho}{E[\tau]\phi_1}\right)^2
\end{align}
where $\psi = s_A^{(1)} \frac{\mathrm{trace}(\mathbf{B})}{q}$.%, and 
%\begin{align}
%\rho_s = \frac{SINR_{out,spatial}}{SINR_{max}}.
%\end{align}

In the small $\sigma^2$ regime this becomes
\begin{align}
E[\rho] \geq 1-\frac{1}{n}.
\end{align}
\end{theorem}
%Since under the Kronecker model the spatial stage projects away the entire clutter subspace (and then some) and spatial Kron STAP is a special case of Kron STAP, this SINR loss bounds the achievable SINR loss of Kron STAP from below. 
Since by \eqref{Eq:ClutterCov} $r \leq q$, the gains of using Kron STAP can be quite significant.

%In practice, however, $\mathbf{d}$ is often not completely orthogonal to the estimated clutter subspace, hence the advantage of using the temporal stage in addition to the spatial stage.

Finally we consider the case where errors occurred in estimating the spatial covariance, either due to subspace estimation error or to $\mathbf{A}$ having a rank greater than one, e.g., due to small calibration errors. Specifically, suppose the estimated (rank one) spatial subspace is $\tilde{\mathbf{h}}$, giving a Kron STAP spatial filter $\mathbf{F}_{spatial} = (\mathbf{I} - \tilde{\mathbf{h}}\tilde{\mathbf{h}}^H)\otimes \mathbf{I}$. Suppose further that spatial filtering of the data is followed by the temporal filter $\mathbf{F}_{temp}$ based on the temporal subspace $\mathbf{U}_B$ estimated from the training data. Define the SINR loss $\rho_t|\tilde{\mathbf{h}}$ from using an estimate of $\mathbf{U}_B$ as
\begin{align}
\rho_t|\tilde{\mathbf{h}} = \frac{\mathrm{SINR}_{out}}{\mathrm{SINR}_{max}(\tilde{\mathbf{h}})}
\end{align}
where $\mathrm{SINR}_{max}(\tilde{\mathbf{h}})$ is the maximum achievable SINR given that the spatial filter is fixed at $\mathbf{F}_{spatial}=(\mathbf{I} - \tilde{\mathbf{h}}\tilde{\mathbf{h}}^H)\otimes \mathbf{I}$. Then it is shown in Appendix \ref{App:Pf} that the expected SINR Loss of the temporal Kron STAP stage is given by the following theorem. %, proved in Appendix \ref{App:SIRV}. 

\begin{theorem}[Kron STAP (temporal stage) SINR]
\label{Thm:Temporal}
%Suppose that a value for the spatial subspace estimate $\tilde{\mathbf{h}}$ and hence $\mathbf{F}_{spatial}$ is fixed. Then for large $n$ and targets with constant Doppler over the integration interval, the SINR loss from using an estimate of $\mathbf{U}_B$ satisfies

Suppose that a value for the spatial subspace estimate $\tilde{\mathbf{h}}$ (with $\|\tilde{\mathbf{h}}\|_2=1$) and hence $\mathbf{F}_{spatial}$ is fixed. Let the steering vector for a constant Doppler target be $\mathbf{d} = \mathbf{d}_A \otimes \mathbf{d}_B$ per \eqref{Eq:Moving}, and suppose that $\mathbf{d}_A$ is fixed and $\mathbf{d}_B$ is arbitrary. Then for large $n$

%\begin{align}
%E[\rho_t | \tilde{\mathbf{h}}] = 1-\frac{1}{n}\sum_{i=1}^{r_b}\left(\frac{E[\tau^2]s_B^{(i)} + \frac{\sigma^2}{\tilde{\mathbf{h}}^H \mathbf{A}\tilde{\mathbf{h}}}}{E[\tau^2]s_B^{(i)}}\right)^2
%\end{align}
\begin{align}
E[\rho_t | &\tilde{\mathbf{h}}] =\\\nonumber& 1-\frac{\kappa}{n}\sum_{i=1}^{r_b}\left(\frac{(E[\tau^2]s_B^{(i)} + \frac{\sigma^2}{\tilde{\mathbf{h}}^H \mathbf{A}\tilde{\mathbf{h}}})(E[\tau^2]s_B^{(i)} + \frac{\sigma^2}{\kappa\tilde{\mathbf{h}}^H \mathbf{A}\tilde{\mathbf{h}}})}{(E[\tau^2]s_B^{(i)})^2}\right)\\\nonumber
&\quad \quad \quad \qquad \kappa = \frac{\tilde{\mathbf{d}}^H_A \mathbf{A}\tilde{\mathbf{d}}_A}{\tilde{\mathbf{h}}^H \mathbf{A}\tilde{\mathbf{h}}}.
\end{align}
%where

In the small $\sigma^2$ regime this becomes
\begin{align}
E[\rho_t | \tilde{\mathbf{h}}] \approx 1-\frac{\kappa r_b}{n}.
\end{align}
\end{theorem}
Note that in the $n \gg p$ regime relevant when $q \gg p$, $\tilde{\mathbf{h}} \approx \mathbf{h}$, where $\mathbf{h}$ is the first singular vector of $\mathbf{A}$. This gives $\tilde{\mathbf{h}}^H \mathbf{A}\tilde{\mathbf{h}} \approx s_A^{(1)}$ and $\kappa \rightarrow 0$ if $\mathbf{A}$ is indeed rank one. Hence, $\kappa$ can be interpreted as quantifying the adverse effect of mismatch between $\mathbf{A}$ and its estimate. %The quantity $\tilde{\mathbf{h}}^H \mathbf{A}\tilde{\mathbf{h}} = s_A^{(1)}$ is dependent 
%REDEFINE THE PHIS SO AS TO COMPARE BETTER WITH SCM.
To avoid cancelation of the moving targets, it is necessary that $r_b \ll q$, and since in the ideal large sample regime all the clutter is removed by the temporal stage, $r_b$ can be smaller than $\mathrm{rank}(\mathbf{B})$. Hence this slower SINR convergence rate in $n$ on a smaller amount of cancelation than the spatial stage (since $\kappa$ should be small) is still faster than that of LR-STAP in general.

%Note that in the $n \gg p$ regime relevant when $q \gg p$, $\tilde{\mathbf{h}} \approx \mathbf{h}$, where $\mathbf{h}$ is the first singular vector of $\mathbf{A}$. This gives $\tilde{\mathbf{h}}^H \mathbf{A}\tilde{\mathbf{h}} \approx s_A^{(1)}$ and $\kappa \rightarrow 0$ if $\mathbf{A}$ is indeed rank one.  %The quantity $\tilde{\mathbf{h}}^H \mathbf{A}\tilde{\mathbf{h}} = s_A^{(1)}$ is dependent 
%%REDEFINE THE PHIS SO AS TO COMPARE BETTER WITH SCM.
%To avoid cancelation of the moving targets, it is necessary that $r_b \ll q$, and since in the ideal large sample regime all the clutter is removed by the temporal stage, $r_b$ can be smaller than $\mathrm{rank}(\mathbf{B})$. Hence this slower convergence rate on a smaller amount of cancelation than the spatial stage (since $\kappa$ should be small) is still faster than that of LR-STAP in general. %Furthermore, if the 
%%DISCUSS GAINS, BOTH MIN AND MAX. QUESTION: CAN WE GET THE GOOD ORDER IF WE PUT AN EIG CONSTRAINT?

%In SAR, it can often be assumed that the noise $\lambda$ is small \cite{ginolhac2014exploiting}. In this regime, we have
%\begin{align}
%\end{align}
%for ??,??, and ?? respectively.

\end{LongerTheorems}

%%%%%%%%%%%%%%%%%%%%%%%%%%%%%%%%%%%%%%%%%%%%%%%%%%%%%%%%%%%%%%%%%%%%%%%%%%%%%%%%%%%%

\begin{ShorterTheorems}

\section{SINR Performance}
\label{Sec:Pred}
%USE CONCENTRATION IN TOP SINGS AS PARAM. FOR ORIGINAL AMT OF CANCEL.
%*Follow the approach of \cite{ginolhac2014exploiting}.
%**Analyzes the effect of noise on subspace estimation error for large n (quadratic perturbation analysis).
%**Converts to a ratio between the achieved SINR and that achievable with infinite training.

For a STAP filter matrix $\mathbf{F}$ and steering vector $\mathbf{d}$, the data filter vector is \eqref{Eq:GenFilt} $\mathbf{w} = \mathbf{F}\mathbf{d}$ \cite{ginolhac2014exploiting}. With a target return of the form $\mathbf{x}_{target} = \alpha \mathbf{d}$, the filter output is given by \eqref{Eq:Breakdown}, and the SINR by \eqref{Eq:SINRdef}.

%\begin{align}
%y = \mathbf{w}^H\mathbf{x} = \alpha \mathbf{w}^H \mathbf{d}+ \mathbf{w}^H \mathbf{n}.
%\end{align}
%MAKE THIS CLEARER!! DEFINE N
%In this paper, the filter $\mathbf{w} = \mathbf{F} \mathbf{d}$ .

%The SINR at the steering vector $\mathbf d$ \cite{ginolhac2014exploiting} is then given by
%\begin{align}
%\mathrm{SINR}_{out} = \frac{|\alpha|^2 |\mathbf{w}^H\mathbf{d}|^2}{E[\mathbf{w}^H \mathbf{n}\mathbf{n}^H \mathbf{w}]} = \frac{|\alpha|^2 |\mathbf{w}^H\mathbf{d}|^2}{\mathbf{w}^H \mathbf{\Sigma} \mathbf{w}}.
%\end{align}
%where $\mathbf{\Sigma}$ is the clutter plus noise covariance in \eqref{Eq:Cov}.

Define $\mathrm{SINR}_{max}$ to be the optimal SINR, achieved at $\mathbf{w}_{opt} = \mathbf{F}_{opt} \mathbf{d}$ \eqref{Eq:opt}.

Suppose that the clutter has covariance of the form \eqref{Eq:KronCov}. %, i.e. it lies in an $r_b$ dimensional subspace, but some ``subspace leakage" occurs. Define ? and ? to be the ratios of ?. 
Assume that the target steering vector $\mathbf{d}$ lies outside both the temporal and spatial clutter subspaces as justified in \cite{ginolhac2014exploiting}. Suppose that LR-STAP is set to use $r$ principal components. Suppose further that Kron STAP uses 1 spatial principal component and $r$ temporal components, so that the total number of principal components of LR-STAP and Kron STAP are equivalent. Under these assumptions, if the noise variance $\sigma^2$ approaches zero the SINR achieved using LR-STAP, Kron STAP or spatial Kron STAP with infinite training samples achieves $\mathrm{SINR}_{max}$ \cite{ginolhac2014exploiting}. %$SINR_{max} = \frac{|\tau|^2}{\sigma^2}$ and achieved at some $\mathbf{w}_{opt}$.
%Similarly, the SINR achieved using KronSTAP with infinite training samples is given by
%\begin{align}
%\end{align}.
%Already a gain. OR IS IT? ONLY IF SUBSPACE LEAKAGE INTO SUBSPACES NOT QUITE ORTHOGONAL TO STEERING VECTOR!!! THEN IT IS. HENCE, SAY EACH STAGE ALONE, UNDER THESE ASSUMPTIONS, IS EQUAL. COMBINING STAGES ONLY HELPS IN AFOREMENTIONED SETTING.

We analyze the asymptotic convergence rates under the finite sample regime. Define the SINR Loss $\rho$ as the loss of performance induced by using the estimated STAP filter $\hat{\mathbf{w}} = \hat{\mathbf{F}}\mathbf{d}$ instead of $\mathbf{w}_{opt}$:
\begin{align}
\rho = \frac{\mathrm{SINR}_{out}}{\mathrm{SINR}_{max}},
\end{align}
where $\mathrm{SINR}_{out}$ is the output signal to interference ratio when using $\hat{\mathbf{w}}$.

%Let $\lambda_i$, $i = 1,\dots,pq$ be the eigenvalues of $\mathbf{\Sigma}_c$. Under the Kronecker model, we have
%\begin{equation}
%{\lambda}_i = \left\{\begin{array}{ll}s_A^{(1)} s_B^{(i)}, & i = 1,\dots, r_b \\ 0 & i > r_b\end{array}\right.
%\end{equation}
%%and $\lambda_i = 0$ for $i > q$, 
%since $\mathbf{A}$ only has one nonzero singular value.

%\begin{theorem}[LR-STAP SINR \cite{ginolhac2014exploiting}]
%\label{Thm:LRSTAP}
%*Degradation of target/clutter SINR (LR-STAP):

It is shown in \cite{ginolhac2014exploiting} that for large $n$ and small $\sigma$, the expected SINR Loss of LR-STAP is
%\begin{align}
%E[\rho] = 1-\frac{1}{n}\sum_{i=1}^r\left(\frac{E[\tau^2]\lambda_i + \sigma^2}{E[\tau^2]\lambda_i}\right)^2,
%\end{align}
%which in the small $\sigma^2$ regime (typical in SAR \cite{ginolhac2014exploiting}) becomes
\begin{align}
E[\rho] = 1-\frac{r}{n}.
\end{align}
%\end{theorem}
%Under the Kronecker model we have
%\begin{align}
%E[\rho] = 1-\frac{1}{n}\sum_{i=1}^{r}\left(\frac{E[\tau^2]s_B^{(i)} + \frac{\sigma^2}{s_A^{(1)}}}{E[\tau^2]s_B^{(i)}}\right)^2.
%\end{align}
%The general case for arbitrary $\sigma$ is found in \cite{ginolhac2014exploiting}.
This approximation is obtained specializing the result in \cite[Prop. 3.1]{ginolhac2014exploiting} to the case of small $\sigma$. 

We now turn to Kron STAP. Note that the Kron STAP filter can be decomposed into a spatial stage (filtering by $\mathbf{F}_{spatial}$) and a temporal stage (filtering by $\mathbf{F}_{temp}$):
\begin{equation}
\mathbf{F}_{KSTAP} = \mathbf{F}_A\otimes \mathbf{F}_B = \mathbf{F}_{spatial} \mathbf{F}_{temp}
\end{equation}
where $\mathbf{F}_{spatial} = \mathbf{F}_A \otimes \mathbf{I}$ and $\mathbf{F}_{temp} = \mathbf{I} \otimes \mathbf{F}_B$ \eqref{Eq:KronSTAP}.
When the clutter covariance fits our model, either the spatial or the temporal stage is sufficient to project away the clutter subspace. The following result assumes the naive estimator 
\begin{align}
\label{Eq:Spat}
\hat{\mathbf{A}} = \mathrm{EIG}_1\left(\frac{1}{q}\sum_{i} \mathbf{S}(i,i) \right) = \hat{\psi}\hat{\mathbf{h}}\hat{\mathbf{h}}^H
\end{align}
for the spatial subspace $\mathbf{h}$ ($\|\mathbf{h}\|_2=1$). %Using this, we have the following theorem.
%This is equivalent to approximating the sample spatial covariance as rank 1. The analysis of \cite{ginolhac2014exploiting} thus applies with $r=1$ and $n' = nq$, except some of the samples are correlated. Using the Kronecker structure of the covariance it is trivial to show (for the SIRV distribution) that the worst case occurs when all the clutter temporal correlations are all $\pm1$, in which case $\frac{1}{q} \sum_i \mathbf{S}(i,i)$ reduces to an $n$ iid sample SCM with Gaussian noise variance $\sigma^2/q$ and we can obtain the following
%\begin{theorem}[Kron STAP SINR]
%\label{Thm:Sp}

For large $n$, small $\sigma$, and using the estimator \eqref{Eq:Spat}, the expected SINR Loss of Kron STAP using the estimator \eqref{Eq:Spat} for the spatial subspace is given by
%\begin{align}
%E[\rho] \geq 1-\frac{1}{n}\left(\frac{E[\tau^2]\psi + \frac{\sigma^2}{q}}{E[\tau^2]\psi}\right)^2 %\leq  \leq 1-\frac{1}{qn}\left(\frac{E[\tau]\phi_1 + \rho}{E[\tau]\phi_1}\right)^2
%\end{align}
%where $\psi = s_A^{(1)} \frac{\mathrm{trace}(\mathbf{B})}{q}$.%, and 
%%\begin{align}
%%\rho_s = \frac{SINR_{out,spatial}}{SINR_{max}}.
%%\end{align}
%
%In the small $\sigma^2$ regime this becomes
\begin{align}
E[\rho] = 1-\frac{1}{n}.
\end{align}
%\end{theorem}
This is a specialization of a result for arbitrary $\sigma$ derived in our technical report \cite[Theorem ??]{greenewald2015kronecker}.
%Since under the Kronecker model the spatial stage projects away the entire clutter subspace (and then some) and spatial Kron STAP is a special case of Kron STAP, this SINR loss bounds the achievable SINR loss of Kron STAP from below. 
Since by \eqref{Eq:ClutterCov} $r \leq q$, the gains of using Kron STAP can be quite significant.

%In practice, however, $\mathbf{d}$ is often not completely orthogonal to the estimated clutter subspace, hence the advantage of using the temporal stage in addition to the spatial stage.

Finally we consider the case where the spatial covariance has estimation errors, either due to subspace estimation error or to $\mathbf{A}$ having a rank greater than one, e.g., due to spatially varying calibration errors. Specifically, suppose the estimated (rank one) spatial subspace is $\tilde{\mathbf{h}}$, giving a Kron STAP spatial filter $\mathbf{F}_{spatial} = (\mathbf{I} - \tilde{\mathbf{h}}\tilde{\mathbf{h}}^H)\otimes \mathbf{I}$. Suppose further that spatial filtering of the data is followed by the temporal filter $\mathbf{F}_{temp}$ based on the temporal subspace $\mathbf{U}_B$ estimated from the training data. Define the SINR loss $\rho_t|\tilde{\mathbf{h}}$ from using an estimate of $\mathbf{U}_B$ as
\begin{align}
\rho_t|\tilde{\mathbf{h}} = \frac{\mathrm{SINR}_{out}}{\mathrm{SINR}_{max}(\tilde{\mathbf{h}})}
\end{align}
where $\mathrm{SINR}_{max}(\tilde{\mathbf{h}})$ is the maximum achievable SINR given that the spatial filter is fixed at $\mathbf{F}_{spatial}=(\mathbf{I} - \tilde{\mathbf{h}}\tilde{\mathbf{h}}^H)\otimes \mathbf{I}$. %, proved in Appendix \ref{App:SIRV}.

%\begin{theorem}[Kron STAP (temporal stage) SINR]
%\label{Thm:Temporal}
We then can obtain the following, which is a specialization of a result for arbitrary $\sigma$ derived in our technical report \cite[Theorem ??]{greenewald2015kronecker}. Suppose that a value for the spatial subspace estimate $\tilde{\mathbf{h}}$ (with $\|\tilde{\mathbf{h}}\|_2=1$) and hence $\mathbf{F}_{spatial}$ is fixed. Let the steering vector for a constant Doppler target be $\mathbf{d} = \mathbf{d}_A \otimes \mathbf{d}_B$ per \eqref{Eq:Moving}, and suppose that $\mathbf{d}_A$ is fixed and $\mathbf{d}_B$ is arbitrary. Then for large $n$ and small $\sigma$, the SINR loss from using an estimate of $\mathbf{U}_B$ follows
%\begin{align}
%E[\rho_t | \tilde{\mathbf{h}}] = 1-\frac{1}{n}\sum_{i=1}^{r_b}\left(\frac{E[\tau^2]s_B^{(i)} + \frac{\sigma^2}{\tilde{\mathbf{h}}^H \mathbf{A}\tilde{\mathbf{h}}}}{E[\tau^2]s_B^{(i)}}\right)^2
%\end{align}
%\begin{align}
%E[\rho_t | &\tilde{\mathbf{h}}] =\\\nonumber& 1-\frac{\kappa}{n}\sum_{i=1}^{r_b}\left(\frac{(E[\tau^2]s_B^{(i)} + \frac{\sigma^2}{\tilde{\mathbf{h}}^H \mathbf{A}\tilde{\mathbf{h}}})(E[\tau^2]s_B^{(i)} + \frac{\sigma^2}{\kappa\tilde{\mathbf{h}}^H \mathbf{A}\tilde{\mathbf{h}}})}{(E[\tau^2]s_B^{(i)})^2}\right)\\\nonumber
%&\quad \quad \quad \qquad \kappa = \frac{\tilde{\mathbf{d}}^H \mathbf{A}\tilde{\mathbf{d}}}{\tilde{\mathbf{h}}^H \mathbf{A}\tilde{\mathbf{h}}}.
%\end{align}
%%where
%
%In the small $\sigma^2$ regime this becomes
\begin{align}
E[\rho_t | \tilde{\mathbf{h}}] \approx 1-\frac{\kappa r_b}{n}, \qquad \kappa = \frac{\tilde{\mathbf{d}}^H_A \mathbf{A}\tilde{\mathbf{d}}_A}{\tilde{\mathbf{h}}^H \mathbf{A}\tilde{\mathbf{h}}}.
\end{align}
where $\tilde{\mathbf{d}}_A =\frac{(\mathbf{I} - \tilde{\mathbf{h}} \tilde{\mathbf{h}}^H)\mathbf{d}_A}{\|(\mathbf{I} - \tilde{\mathbf{h}} \tilde{\mathbf{h}}^H)\mathbf{d}_A\|_2}$.
%Then it is shown in the technical report \cite{??} that the expected SINR Loss of the temporal Kron STAP stage is given by the following theorem. The case of arbitrary $\sigma$ can also be found in the report \cite{??}. 

Note that in the $n \gg p$ regime relevant when $q \gg p$, $\tilde{\mathbf{h}} \approx \mathbf{h}$, where $\mathbf{h}$ is the first singular vector of $\mathbf{A}$. This gives $\tilde{\mathbf{h}}^H \mathbf{A}\tilde{\mathbf{h}} \approx s_A^{(1)}$ and $\kappa \rightarrow 0$ if $\mathbf{A}$ is indeed rank one. Hence, $\kappa$ can be interpreted as quantifying the adverse effect of mismatch between $\mathbf{A}$ and its estimate. %The quantity $\tilde{\mathbf{h}}^H \mathbf{A}\tilde{\mathbf{h}} = s_A^{(1)}$ is dependent 
%REDEFINE THE PHIS SO AS TO COMPARE BETTER WITH SCM.
To avoid cancelation of the moving targets, it is necessary that $r_b \ll q$, and since in the ideal large sample regime all the clutter is removed by the temporal stage, $r_b$ can be smaller than $\mathrm{rank}(\mathbf{B})$. Hence this slower SINR convergence rate in $n$ on a smaller amount of cancelation than the spatial stage (since $\kappa$ should be small) is still faster than that of LR-STAP in general. %Furthermore, if the 
%DISCUSS GAINS, BOTH MIN AND MAX. QUESTION: CAN WE GET THE GOOD ORDER IF WE PUT AN EIG CONSTRAINT?

\end{ShorterTheorems}

\section{Numerical Results}
\label{Sec:Results}
\subsection{Dataset}

%For evaluation of our detection methods, we use two multichannel radar datasets: one synthetic and one real dataset.

%The KASSPER dataset is a physics-based synthetic multi-antenna radar dataset. The synthesized scene consists of a region of western California with several hundred moving ground targets on the local road network. The antenna array is an 11 element linear array, and only ?? pulses are available. Additional details of the simulation can be found in ????. Since the true target locations are available, ROC curves can be computed for the tested detection strategies.

For evaluation of the proposed Kron STAP methods, we use measured data from the 2006 Gotcha SAR GMTI sensor collection \cite{GotchaData}. This dataset consists of SAR passes through a circular path around a small scene containing various moving and stationary civilian vehicles. The example images shown in the figures are formed using the backprojection algorithm with Blackman-Harris windowing as in \cite{newstadt2013moving}. For our experiments, we use 31 seconds of data, divided into 1 second (2171 pulse) coherent integration intervals.

As there is no ground truth for all targets in the Gotcha imagery, target detection performance cannot be objectively quantified by ROC curves. %GPS data exists for some targets but not all, so we resorted to truthing the dataset by comparing and analyzing the results of the best detection methods available. 
We rely on non ROC measures of performance for the measured data, and use synthetically generated data to show ROC performance gains. In several experiments we do make reference to several higher amplitude example targets in the Gotcha dataset. These were selected by comparing and analyzing the results of the best detection methods available.

\subsection{Simulations}
\begin{figure}[htb]
\centering
\includegraphics[width=2.8in]{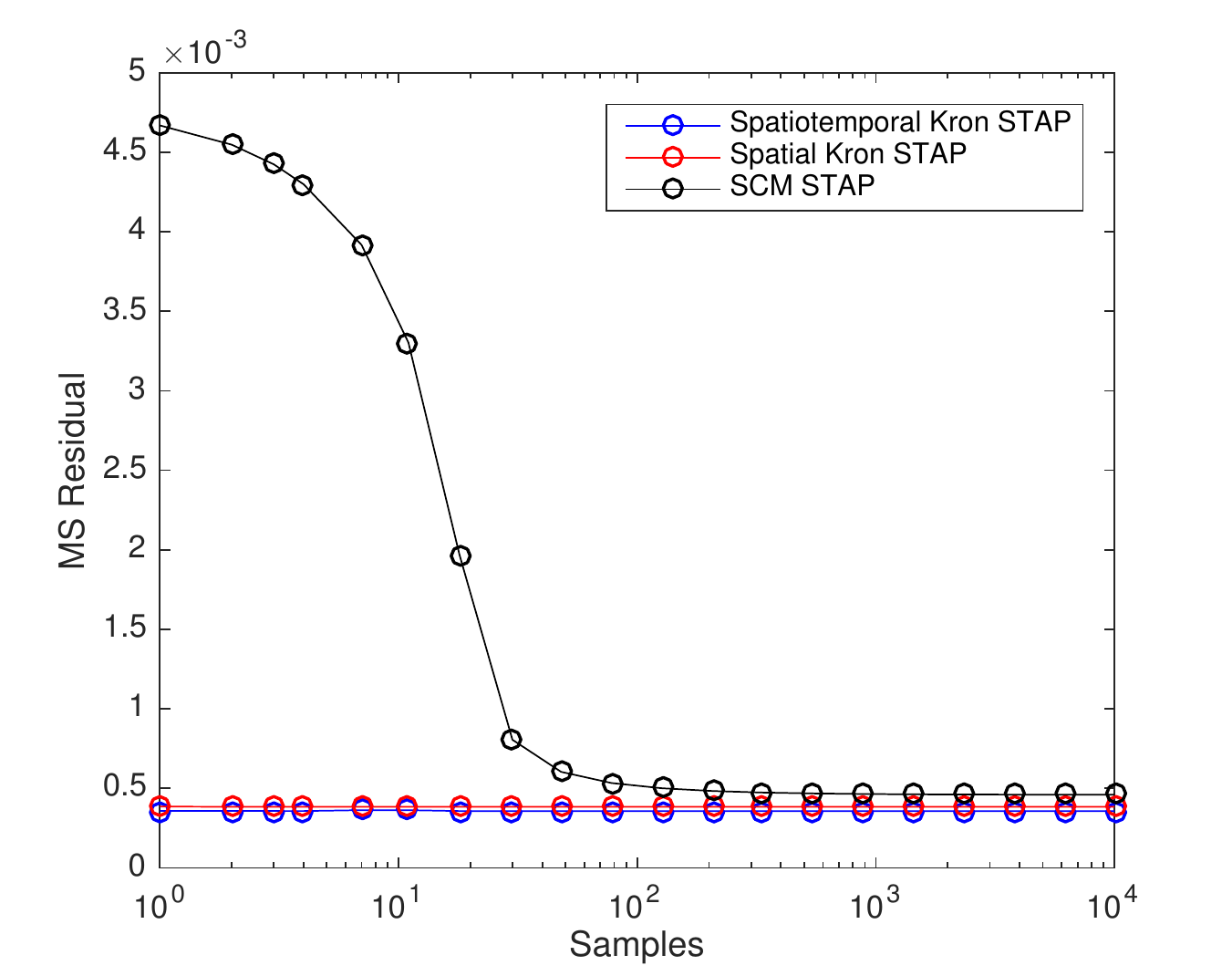}\includegraphics[width=2.8in]{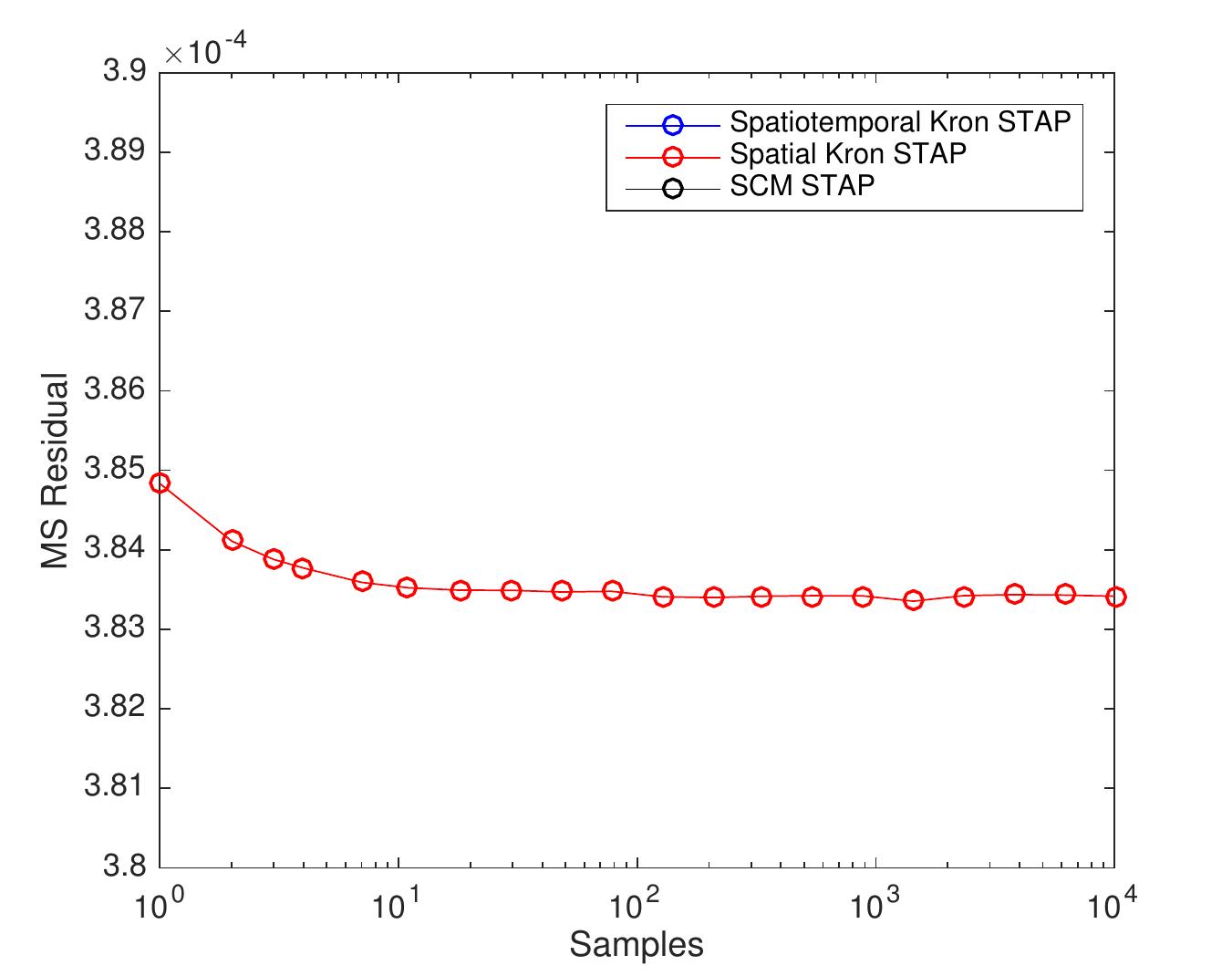}
%\end{centering}
\caption{Average mean squared residual (MSR), as a function of the number of training samples, of noisy synthetic clutter filtered by spatio-temporal Kron STAP, spatial only Kron STAP, and unstructured LR-STAP (SCM STAP) filters. On the right a zoomed in view of a Kron STAP curve is shown. Note the rapid convergence and low MSE of the Kronecker methods.}
\label{Fig:MS}
\end{figure}

We generated synthetic clutter plus additive noise samples having a low rank Kronecker product covariance. The covariance we use to generate the synthetic clutter via the SIRV model was learned from a set of example range bins extracted from the Gotcha dataset, letting the SIRV scale parameter $\tau^2$ in \eqref{Eq:Cov} follow a chi-square distribution. We use $p=3$, $q=150$, $r_b= 20$, and $r_a = 1$, and generate both $n$ training samples and a set of testing samples. The rank of the left Kronecker factor $\mathbf A$, $r_a$, is 1 as dictated by the spatially invariant antenna calibration assumption and we chose $r_b = 20$ based on a scree plot, i.e., $20$ was the location of  the knee of the spectrum of $\mathbf{B}$. Spatio-temporal Kron-STAP, Spatial-only Kron-STAP, and LR-STAP were then used to learn clutter cancelation filters from the training clutter data. The learned filters were then applied to testing clutter data, the mean squared value (MS Residual) of the resulting residual (i.e. $(1/M)\sum_{m=1}^M \|\mathbf{F} \mathbf{x}_m \|_2^2$) was computed, and the result is shown in Figure \ref{Fig:MS} as a function of $n$. The results illustrate the much slower convergence rate of unstructured LR-STAP. as compared to the proposed Kron STAP, which converges after $n=1$ sample. The mean squared residual does not go to zero with increasing training sample size because of the additive noise floor.

\begin{figure}[htb]
\begin{centering}
\includegraphics[width=2.8in]{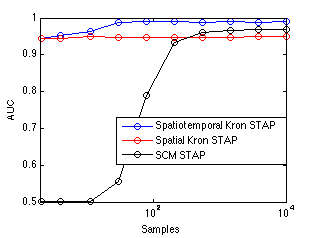}%\includegraphics[width=1.8in]{ROC_true_new.png}

\end{centering}
\caption{Area-under-the-curve (AUC) for the ROC associated with detecting a synthetic target using the steering vector with the largest return, when slight spatial nonidealities exist in the true clutter covariance. %Right: For comparison, the AUCs for detecting a weaker target when the target's true steering vector (i.e. cross range position and velocity) is known. 
Note the rapid convergence of the Kronecker methods as a function of the number of training samples, and the superior performance of spatio-temporal Kron STAP to spatial-only Kron STAP when the target's steering vector $\mathbf{d}$ is unknown. }
\label{Fig:ROC}
\end{figure}

To explore the effect of model mismatch due to spatially variant antenna calibration errors ($r_a>1$), we simulated data with a clutter spatial covariance $\mathbf{A}$ having rank 2 with non-zero eigenvalues equal to 1 and $1/30^2$. The STAP algorithms remain the same with $r_a = 1$,  %(otherwise only a rank one potential target subspace would remain)
and synthetic range bins containing both clutter and a moving target are used in testing the effect of this model mismatch on the STAP algorithms. The STAP filter response, maximized over all possible steering vectors, is used as the detection statistic. The AUC of the associated ROC curves is plotted on the left in Figure \ref{Fig:ROC} as a function of the number of training samples. Note again the poor performance and slow convergence of LR-STAP, and that spatio-temporal Kron-STAP converges very quickly to the optimal spatial Kron-STAP performance, and more slowly converges to a superior performance as the temporal filter estimate converges. 

%A similar plot for the case of a weaker target with the filtering response from the known target steering vector used as the detection statistic is shown on the right side of Figure \ref{Fig:ROC}. No gains are observed for spatio-temporal Kron-STAP over spatial Kron-STAP in this case because ???.

\begin{figure}[htb]
%\begin{centering}
\centering
\includegraphics[width=2.8in]{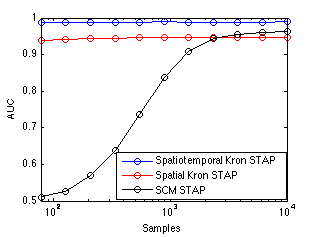}
%\end{centering}
\caption{Robustness to corrupted training data: AUCs for detecting a synthetic target using the maximum steering vector when (in addition to the spatial nonidealities) 5\% of the training range bins contain targets with random location and velocity in addition to clutter. Note that relative to Figure \ref{Fig:ROC} %(left) 
LR-STAP has degraded significantly, whereas the Kronecker methods have not.}
\label{Fig:ROCC}
\end{figure}

Finally, we repeat the AUC vs. sample complexity experiment of the previous paragraph with 5\% of the training data having synthetic moving targets with random Doppler shifts. The results are shown in Figure \ref{Fig:ROCC}. As predicted by the theory in Subsection \ref{Sec:Robust}, the Kronecker methods remain largely unaffected by the presence of corrupting targets in the training data, whereas significant losses are sustained by LR-STAP. This confirms the superior robustness of the proposed Kronecker structured covariance in our Kron STAP method.

%Example clutter covariance matrix estimates are shown in Figure \ref{Fig:Cov}.
%\begin{figure}[htb]
%\begin{centering}
%\includegraphics[width=3in]{Covs.png}
%\end{centering}
%\caption{Left: 150 range bin estimate. Right: 50 range bin estimate.}
%\label{Fig:Cov}
%\end{figure}

%Hence, we resort to evaluating our results anecdotally and comparing to other multichannel detection algorithm detections.

%*Public release Gotcha GMTI challenge dataset. $p=3$ antenna channels; circular SAR; vast number of pulses. Used ~1 second integration time.

%TABLE OF RADAR PARAMETERS. %ALSO REF, WHATEVER GREG DID.

\subsection{Gotcha Experimental Data}
%\subsubsection{KASSPER}
%TRY APPROXIMATING THE GIVEN CLUTTER COVARIANCES USING THE KRONECKER MODEL. TO EVALUATE BIAS.

%TABLE OF RUNTIMES!!!! SEPARATE ONES FOR LEARNING AND INFERENCE

%MOTIVATE USE OF IMAGE INSTEAD OF PULSES.

In this subsection, STAP is applied to the Gotcha dataset. For each range bin we construct steering vectors $\mathbf{d}_i$ corresponding to 150 cross range pixels. In single antenna SAR imagery, each cross range pixel is a Doppler frequency bin that corresponds to the cross range location for a stationary target visible at that SAR Doppler frequency, possibly complemented by a moving target that appears in the same bin. Let $\mathbf{D}$ be the matrix of steering vectors for all 150 Doppler (cross range) bins in each range bin. Then the SAR images at each antenna are given by $\tilde{\mathbf{x}} = \mathbf{I}\otimes\mathbf{D}^H\mathbf{x}$ and the STAP output for a spatial steering vector $\mathbf{h}$ and temporal steering $\mathbf{d}_i$ (separable as noted in \eqref{Eq:Moving}) is the scalar
\begin{align}
y_i(\mathbf{h}) = (\mathbf{h}\otimes \mathbf{d}_i)^H \mathbf{F} \mathbf{x}%\\\nonumber
%= (\mathbf{\alpha}\otimes \mathbf{d}_i)^H (\mathbf{I}\otimes \mathbf{D}\mathbf{D}^H) \mathbf{F} \mathbf{x}\\\nonumber
%=
\end{align}
Due to their high dimensionality, plots for all values of $\mathbf{h}$ and $i$ cannot be shown. Hence, for interpretability we produce images where for each range bin the $i$th pixel is set as $\max_{\mathbf{h}} |y_i(\mathbf{h})|$. More sophisticated detection techniques could invoke priors on $\mathbf{h}$, but we leave this for future work.

Shown in Figure \ref{Fig:Examples} are results for several examplar SAR frames, showing for each example the original SAR (single antenna) image, the results of spatio-temporal Kronecker STAP, the results of Kronecker STAP with spatial filter only, the amount of enhancement (smoothed dB difference between STAP image and original) at each pixel of the spatial only Kronecker STAP, standard unstructured STAP with $r=25$ (similar rank to Kronecker covariance estimate), and standard unstructured STAP with $r = 40$. Note the significantly improved contrast of Kronecker STAP relative to the unstructured methods between moving targets (high amplitude moving targets marked in red in the figure) and the background. Additionally, note that both spatial and temporal filtering achieve significant gains. Due to the lower dimensionality, LR-STAP achieves its best performance for the image with fewer pulses, but still remains inferior to the Kronecker methods. %However, the inclusion of additional pulses seems to improve target contrast for the Kronecker methods, and the additional smearing can be exploited using techniques such as shear averaging \cite{fienup2001detecting} which we do not consider here.

%\subsubsection{Gotcha}

%https://restricted.vdl.afrl.af.mil/webdav/programs/atrpedia/Nontechnical_Materials/People/Greenewald_Kristjan/2014PageFigs/Pic6.png

%EXAMPLES Giant figure of examples. Include focused, and partial. Half page?

%MSE

\begin{figure}[htb]
\centering
\includegraphics[width=2.8in]{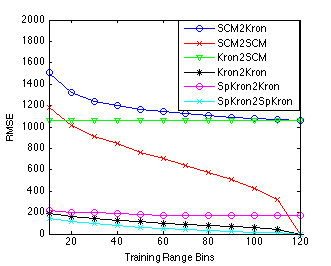}\includegraphics[width=2.8in]{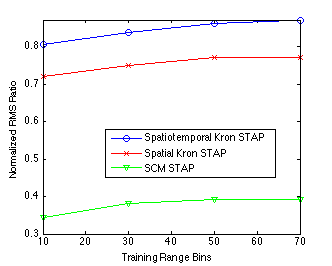}
%\end{centering}
\caption{Gotcha dataset. Left: Average RMSE of the output of the Kronecker, spatial only Kronecker, and unstructured STAP filters relative to each method's maximum training sample output. Note the rapid convergence and low RMSE of the Kronecker methods. Right: Normalized ratio of the RMS magnitude of the brightest pixels in each target relative to the RMS value of the background, for the output of each of Kronecker STAP, spatial Kronecker STAP, and unstructured STAP.}
\label{Fig:RMSE}
\end{figure}

To analyze convergence behavior, a Monte Carlo simulation was conducted where random subsets of the (bright object free) available training set were used to learn the covariance and the corresponding STAP filters. The filters were then used on each of the 31 1-second SAR imaging intervals and the MSE between the results and the STAP results learned using the entire training set were computed (Figure \ref{Fig:RMSE}). %The results are shown in Figure \ref{Fig:RMSE}. 
Note the rapid convergence of the Kronecker methods relative to the SCM based method, as expected. %This confirms that significantly fewer training samples are required under the proposed Kronecker clutter covariance model.

%RATIO OF POWERS
Figure \ref{Fig:RMSE} (right) shows the normalized ratio of the RMS magnitude of the 10 brightest filter outputs $y_i(\mathbf{h})$ for each ground truthed target to the RMS value of the background, computed for each of the STAP methods as a function of the number of training samples. This measure is large when the contrast of the target to the background is high. The Kronecker methods clearly outperform LR-STAP.

\begin{CD3}
\subsection{Multipass Kron STAP}
%EXAMPLES. %MSE RATIO PLOTS. WOULD NEED TO MARK OTHER PASS THOUGH.
Representative two pass Kronecker STAP results are shown in Figure \ref{Fig:CD}, comparing to two pass LR-STAP and to standard (gain calibrated) incoherent change detection. For the STAP methods, noncoherent change detection is performed following filtering by reforming each image (via maximum steering vectors as in the previous section) and subtracting the resulting pixel magnitudes. It can be seen that additional clutter cancelation capabilities can be gained by using Kronecker STAP on multiple passes.

As in the single pass case, Figure \ref{Fig:RMSE_chng} shows relative RMSE convergence results and the normalized RMS ratio between targets and background. Again, Kron STAP outperforms the other methods, and both STAP methods outperform standard incoherent change detection.

\begin{figure}[htb]
\centering
\includegraphics[width=4in]{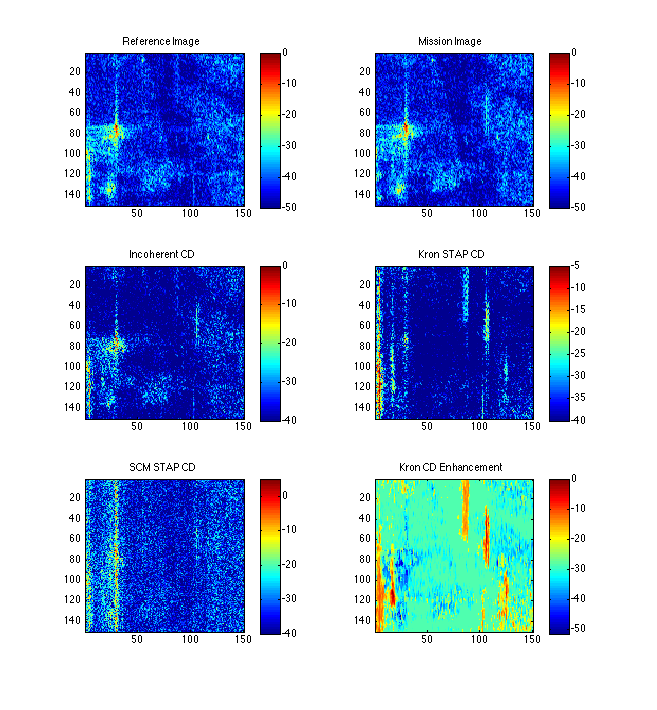}
\caption{Multipass STAP. An example reference and mission image pair are shown, both of which include moving targets. Shown are the results of incoherent change detection, multipass spatio-temporal Kron STAP, multipass LR-STAP, and the multipass spatial Kron STAP enhancement. Note the superior moving target enhancement of the Kronecker methods.}
\label{Fig:CD}
\end{figure}

\begin{figure}[htb]
\centering
\includegraphics[width=2.1in]{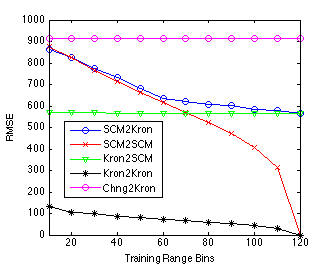}\includegraphics[width=2.1in]{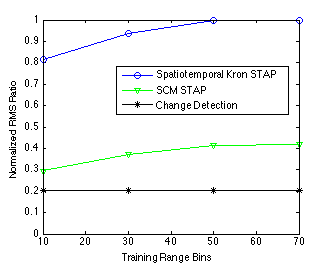}
\caption{Left: Average RMSE of the output of the Kronecker and unstructured STAP filters and incoherent change detection relative to each method's maximum training sample output. Note the rapid convergence and low RMSE of the Kronecker methods. Right: Normalized ratio of the RMS magnitude of the brightest pixels in each target relative to the RMS value of the background, for the output of each of Kronecker STAP, incoherent change detection, and unstructured STAP.}
\label{Fig:RMSE_chng}
\end{figure}
\end{CD3}

%\begin{algorithm}[H]
%\caption{Proximal Gradient Robust KronPCA}
%\label{alg:SVT}
%\begin{algorithmic}[1]

%\STATE $\mathbf{B} = \mathbf{P}\mathcal{R}(\hat{\mathbf{\Sigma}}_{SCM})$
%\STATE Initialize $\mathbf{M},\mathbf{S},\mathbf{L}$.
%\WHILE{Not converged}
%\STATE $\mathbf{L}^k = \mathbf{SVT}_{\tau_k\lambda_\Theta'}(\mathbf{M}^{k-1}-\mathbf{S}^{k-1})$
%\FOR{ $j \in \mathcal{I}$}
%\STATE $\mathbf{S}^k_{j+p_t} = \mathbf{soft}_{\tau_k\lambda_\Gamma'c_j}(\mathbf{M}^{k-1}_{j+p_t}-\mathbf{L}^{k-1}_{j+p_t})$
%\ENDFOR
%\STATE $\mathbf{M}^k = \mathbf{L}^k + \mathbf{S}^k - \tau_k(\mathbf{L}^k + \mathbf{S}^k-\mathbf{B})$
%\ENDWHILE
%\STATE $\hat{\mathbf{\Sigma}} = \mathcal R^{-1}\left(\mathbf{P}^{T} \left( {\mathbf{L}}+{\mathbf{S}} \right)\right)$
%\RETURN $\hat{\mathbf{\Sigma}}$
%\end{algorithmic}
%\end{algorithm}

%It should be noted that omitting the Toeplitz constraint is trivial and achieved by omitting the $\mathcal{P}$ operator steps.

\section{Conclusion}
\label{Sec:Conclusion}
In this paper, we proposed a new method for clutter rejection in high resolution multiple antenna synthetic aperture radar systems with the objective of detecting moving targets. Stationary clutter signals in multichannel single-pass radar were shown to have Kronecker product structure where the spatial factor is rank one and the temporal factor is low rank. Exploitation of this structure was achieved using the Low Rank KronPCA covariance estimation algorithm, and a new clutter cancelation filter exploiting the space-time separability of the covariance was proposed. The resulting clutter covariance estimates were applied to STAP clutter cancelation, exhibiting significant detection performance gains relative to existing low rank covariance estimation techniques. As compared to standard unstructured low rank STAP methods, the proposed Kronecker STAP method reduces the number of required training samples and enhances the robustness to corrupted training data. These performance gains were analytically characterized using a SIRV based analysis and experimentally confirmed using simulations and the Gotcha SAR GMTI dataset.

\begin{appendices}
\section{Derivation of Algorithm \ref{alg:LRKron}}
\label{App:LRKron}
We have the following objective function:
\begin{equation}
\label{Eq:SparseOptApp}
\min_{\mathrm{rank}({\mathbf{A}}) = r_a,\mathrm{rank}({\mathbf{B}}) = r_b}\| \mathbf{S}-{ \mathbf{A}}\otimes{\mathbf{B}}\|_F^2.
\end{equation}
%Let $\mathbf{S} = \mathbf{\Sigma}_{SCM}$.

To derive the alternating minimization algorithm, fix $\mathbf{B}$ (symmetric) and minimize \eqref{Eq:SparseOptApp} over low rank $\mathbf{A}$:
\begin{align}
\label{Eq:EY}
\arg&\min_{\mathrm{rank}({\mathbf{A}}) = r_a}\| \mathbf{S}-{ \mathbf{A}}\otimes{\mathbf{B}}\|_F^2\nonumber\\\nonumber
=&\arg\min_{\mathrm{rank}({\mathbf{A}}) = r_a}\sum_{i,j}^q \| \mathbf{S}(i,j)-b_{ij}{ \mathbf{A}}\|_F^2\\\nonumber
=&\arg\min_{\mathrm{rank}({\mathbf{A}}) = r_a}\sum_{i,j}^q |b_{ij}|^2\|\mathbf{A}\|_F^2 - 2 \mathrm{Re}[b_{ij} \left\langle \mathbf{A},\mathbf{S}^*(i,j)\right\rangle]\\\nonumber
=&\arg\min_{\mathrm{rank}({\mathbf{A}}) = r_a} \|\mathbf{A}\|_F^2 - 2 \mathrm{Re}\left[\left\langle \mathbf{A},\frac{\sum_{i,j}^q b_{ij}\mathbf{S}^*(i,j)}{\|\mathbf{B}\|_F^2}\right\rangle\right]\\
=&\arg\min_{\mathrm{rank}({\mathbf{A}}) = r_a} \left\|\mathbf{A} - \frac{\sum_{i,j}^q b^*_{ij}\mathbf{S}(i,j)}{\|\mathbf{B}\|_F^2}\right\|_F^2
\end{align}
where $b_{ij}$ is the $i,j$th element of $\hat{\mathbf{B}}$ and $b^*$ denotes the complex conjugate of $b$. This last minimization problem \eqref{Eq:EY} can be solved by the SVD via the Eckart-Young theorem \cite{eckart1936approximation}. First define
\begin{align}
\mathbf{R}_A= \frac{\sum_{i,j}^q b^*_{ij}\mathbf{S}(i,j)}{\|\mathbf{B}\|_F^2},
\end{align}
and let $\mathbf{u}_i^{A},\sigma_i^{A}$ be the eigendecomposition of $\mathbf{R}_A$. The eigenvalues are real and positive because $\mathbf{R}_A$ is positive semidefinite (psd) Hermitian if $\mathbf{B}$ is psd Hermitian \cite{werner2008estimation}. Hence by Eckardt-Young the minimizer of the objective \eqref{Eq:EY} is 
\begin{align}
\hat{\mathbf{A}}(\mathbf{B})  =\mathrm{EIG}_{r_a}(\mathbf{R}_A)= \sum_{i = 1}^{r_a} \sigma_i\mathbf{u}_i ^{A} (\mathbf{u}_i^{A})^H.
\end{align}
Similarly, minimizing \eqref{Eq:SparseOptApp} over $\mathbf{B}$ with fixed positive semidefinite Hermitian $\mathbf{A}$ gives
\begin{align}
\label{Eq:BofA}
\hat{\mathbf{B}}(\mathbf{A})  =\mathrm{EIG}_{r_b}(\mathbf{R}_B) =  \sum_{i = 1}^{r_b} \sigma_i^{B}\mathbf{u}_i^{B} (\mathbf{u}_i^{B})^H,
\end{align}
where now $\mathbf{u}_i^{B}, \sigma_i^{B}$ describes the eigendecomposition of 
\begin{align}
\label{Eq:RB}
\mathbf{R}_B= \frac{\sum_{i,j}^p a^*_{ij}\bar{\mathbf{S}}(i,j)}{\|\mathbf{A}\|_F^2}.
\end{align}
Iterating between computing $\hat{\mathbf{A}}(\mathbf{B})$ and $\hat{\mathbf{B}}(\mathbf{A})$ completes the alternating minimization algorithm. 

By induction, initializing with either a psd Hermitian $\mathbf{A}$ or $\mathbf{B}$ and iterating until convergence will result in an estimate $\hat{\mathbf{A}}\otimes \hat{\mathbf{B}}$ of the covariance that is psd Hermitian since the set of positive semidefinite Hermitian matrices is closed. 
%\section{Proof of Theorem }

%%%%%%%
\begin{ThmApp}
\section{Proof of Theorem \ref{Thm:Temporal}}
\label{App:Pf}
After the spatial stage of Kron STAP projects away \eqref{Eq:SPKR} the estimated spatial subspace $\tilde{\mathbf{h}}$ (where $\|\tilde{\mathbf{h}}\|_2=1$) the remaining clutter has a covariance given by
\begin{align}
((\mathbf{I} - \tilde{\mathbf{h}}\tilde{\mathbf{h}}^H)\mathbf{A}(\mathbf{I} - \tilde{\mathbf{h}}\tilde{\mathbf{h}}^H))\otimes \mathbf{B}.
\end{align}

By \eqref{Eq:Moving}, the steering vector for a (constant Doppler) moving target is of the form $\mathbf{d} = \mathbf{d}_A \otimes \mathbf{d}_B$. Hence, the filtered output is 
\begin{align}
y &= \mathbf{w}^H \mathbf{x} = \mathbf{d}^H\mathbf{F} \mathbf{x}\\\nonumber
& = (\mathbf{d}_A^H \otimes \mathbf{d}_B^H) (\mathbf{F}_A \otimes \mathbf{F}_B)\mathbf{x} \\\nonumber &= ((\mathbf{d}_A^H \mathbf{F}_A)\otimes (\mathbf{d}_B^H \mathbf{F}_B)) \mathbf{x} \\\nonumber &= \mathbf{d}_B^H \mathbf{F}_B \left(\left(\mathbf{d}_A^H \left(\mathbf{I} - \tilde{\mathbf{h}} \tilde{\mathbf{h}}^H\right)\right)\otimes \mathbf{I}\right)\mathbf{x}
\end{align}
Let $\tilde{\mathbf{d}}_A =(\mathbf{I} - \tilde{\mathbf{h}} \tilde{\mathbf{h}}^H)\mathbf{d}_A$  and define $\tilde{\mathbf{c}} = \left(\tilde{\mathbf{d}}_A^H \otimes \mathbf{I}\right)\mathbf{c}$. Then 
\begin{align}
y = \mathbf{d}_B^H \mathbf{F}_B (\tau \tilde{\mathbf{c}} + \tilde{\mathbf{n}}),
\end{align}
where $\tilde{\mathbf{n}} = (\tilde{\mathbf{d}}_A \otimes \mathbf{I})\mathbf{n}$ and
\begin{align}
\mathrm{Cov}[\tilde{\mathbf{c}}] =& (\tilde{\mathbf{d}}_A^H \mathbf{A} \tilde{\mathbf{d}}_A) \mathbf{B} \\\nonumber
\mathrm{Cov}[\tilde{\mathbf{n}}] =& \sigma^2 \mathbf{I},
\end{align}
which are proportional to $\mathbf{B}$ and $\mathbf{I}$ respectively. The scalar $(\tilde{\mathbf{d}}_A^H \mathbf{A} \tilde{\mathbf{d}}_A)$ is small if $\mathbf{A}$ is accurately estimated, hence improving the SINR but not affecting the SINR loss.
Thus, the temporal stage of Kron STAP is equivalent to single channel LR-STAP with clutter covariance $(\tilde{\mathbf{d}}_A^H \mathbf{A} \tilde{\mathbf{d}}_A)\mathbf{B}$ and noise variance $\sigma^2$.

%EQUATIONS CREATING FINAL OUTPUT AS BEING SEPARABLY PROCESSED. 
%GOAL: SHOW NOISE ENDURES, AND EQUIVALENT TO JUST DOING TEMPORAL PROCESSING. WITH A LOSS DUE TO ERRONEOUS INNER PRODUCT. 
%WHICH SHOULD BE BOUNDED. IF POSSIBLE.

%The spatial factor of this expression will ideally vanish for large $n$.
%Suppose $\mathbf{B}$ has singular values given by $\phi^{B}_i$ for $i =1,\dots, q$. 

%The projections of the noise onto each of the principal components remains bounded by $\lambda$.

%IN DERIV, MUST MODIFY SIGMA FOR PREVIOUS PROJECTION. OR NOT????

%Then we can assume that the rank one $\mathbf{A}$ estimate is fixed (up to a scaling constant) with small error. Specifically, let
%\begin{align}
%Put L2 norm bound on estimation error. Assert happens whp?
%\end{align}

Given a fixed $\hat{\mathbf A} = \tilde{\mathbf{h}}\tilde{\mathbf{h}}^H$, Algorithm \ref{alg:LRKron} dictates \eqref{Eq:RB}, \eqref{Eq:BofA} that 
\begin{align}
\mathbf{R}_B= \sum_{i,j}^p \tilde{h}^*_{i}\tilde{h}^*_{j}\bar{\mathbf{S}}(i,j)\\\nonumber
\hat{\mathbf{B}} = \mathrm{EIG}_{r_b}(\mathbf{R}_B),
\end{align}
which is thus the low rank approximation of the sample covariance of 
\begin{align}
\mathbf{x}_h = \mathbf{x}_{c,h} + \mathbf{n}_h =  (\tilde{\mathbf{h}}\otimes \mathbf I)^H (\mathbf{x}_c + \mathbf{n}).
\end{align}
%Let the spectrum of $\mathbf{B}$ be given by $\phi^B_i$, $i= 1,\dots q$. 
Since $\mathbf{x}_c = \tau \mathbf{c}$, $\mathbf{x}_{c,h} = \tau (\tilde{\mathbf{h}}\otimes \mathbf I)^H \mathbf{c}$ is an SIRV (Gaussian random vector $(\tilde{\mathbf{h}}\otimes \mathbf I)^H \mathbf{c}$ scaled by $\tau$) with 
\begin{equation}
\mathrm{Cov}[\mathbf{x}_{c,h}] = \tau^2 (\tilde{\mathbf{h}}^H \mathbf{A} \tilde{\mathbf{h}})\mathbf{B}
\end{equation}
%and $\mathbf{x}_{c,h}$ is an SIRV. 
Furthermore, $\mathbf{n}_{h} =(\tilde{\mathbf{h}}\otimes \mathbf I)^H \mathbf{n}$ which is Gaussian with covariance $\sigma^2 \mathbf{I}$. Thus, in both training and filtering the temporal stage of Kron STAP is exactly equivalent to single channel LR STAP. Thus to prove Theorem \ref{Thm:Temporal} it is straightforward to apply the analysis in the proof of Theorem \ref{Thm:LRSTAP} \cite{ginolhac2014exploiting}, with the noise variance in training effectively being $\kappa = \frac{\tilde{\mathbf{d}}^H \mathbf{A}\tilde{\mathbf{d}}}{\tilde{\mathbf{h}}^H \mathbf{A}\tilde{\mathbf{h}}}$ times the noise variance in testing.

\end{ThmApp}

\end{appendices}
\begin{figure*}[]
\begin{centering}
\includegraphics[width=6.6in]{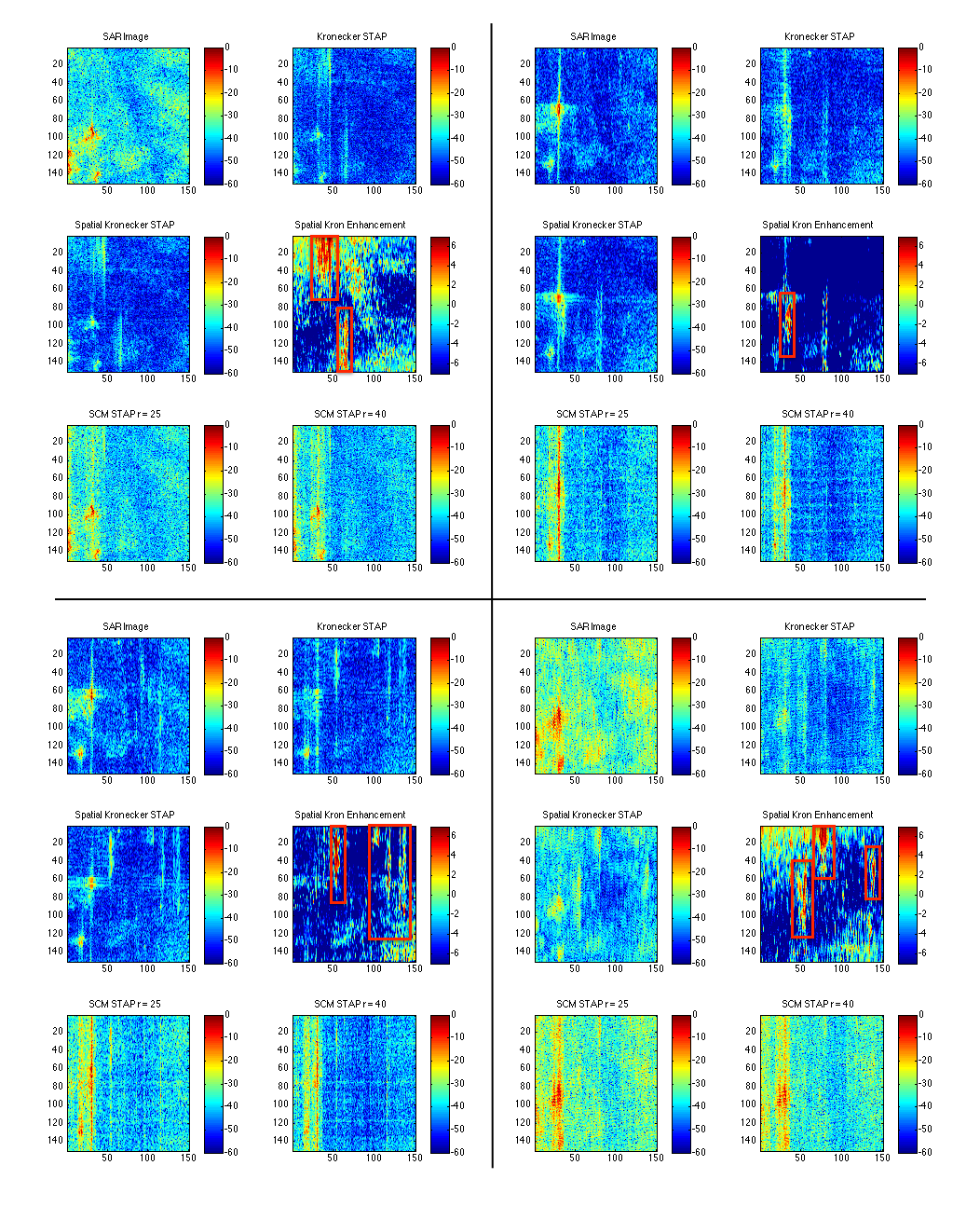}%7.25
\end{centering}
\caption{Four example radar images from the Gotcha dataset along with associated STAP results. The lower right example uses 526 pulses, the remaining three use 2171 pulses. Several moving targets are highlighted in red in the spatial Kronecker enhancement plots. Note the superiority of the Kronecker methods. Used Gotcha dataset ``mission" pass, starting times: upper left, 53 sec.; upper right, 69 sec.; lower left, 72 sec.; lower right 57.25 sec.} %MAYBE VERTICAL AND HORIZONTAL LINES, OR COLOR CODED FILL. LABEL THE AXES, IDIOT. MAYBE ALSO SHOW FACTORS SPECTRA? DRAW ARROWS AND SPATIAL THRESHOLDS TO CONNECT.}
\label{Fig:Examples}
\end{figure*}

\bibliographystyle{IEEETranS}
\bibliography{CAMSAP_bib}

\end{document}